\documentclass[12pt]{article}

\usepackage{amsmath,amssymb,graphicx} %use drftcite in draftmode otherwise
\usepackage{epsf}
\newcommand{\feyn}[4]{\hline $#1$ & $#2$ & $#3$ & $#4$ \\}
\newcommand{\dslash}[1]{\displaystyle{\not} #1}
\newcommand{\beq}{\begin{eqnarray}}% can be used as {equation} or {eqnarray}
\newcommand{\eeq}{\end{eqnarray}}
\newcommand{\arline}{\nonumber \\}
\newcommand{\centeron}[2]{{\setbox0=\hbox{#1}\setbox1=\hbox{#2}\ifdim
                           \wd1>\wd0\kern.5\wd1\kern-.5\wd0\fi \copy0
                           \kern-.5\wd0\kern-.5\wd1\copy1\ifdim\wd0>\wd1
                           \kern.5\wd0\kern-.5\wd1\fi}}
\newcommand{\ltap}{\>\centeron{\raise.35ex\hbox{$<$}}
                   {\lower.65ex\hbox{$\sim$}}\>}
\newcommand{\gtap}{\>\centeron{\raise.35ex\hbox{$>$}}
                   {\lower.65ex\hbox{$\sim$}}\>}
\newcommand{\gsim}{\mathrel{\gtap}}
\newcommand{\lsim}{\mathrel{\ltap}}

\newcommand\ZZ{\hbox{\zfont Z\kern-.4emZ}}
\font\zfont = cmss10 %scaled \magstep1

\newcommand{\iden}{{\mathbf 1}}

\def\beqa{\begin{eqnarray}}
\def\eeqa#1{\label{#1}\end{eqnarray}}
\def\eeqan{\end{eqnarray}}
\def\EPS2{\left( \frac{v}{f} \right)^2}

\def\be{\begin{equation}}
\def\ee{\end{equation}}
\def\bea{\begin{eqnarray}}
\def\eea{\end{eqnarray}}
\def\lsim{\mathrel{\raise.3ex\hbox{$<$\kern-.75em\lower1ex\hbox{$\sim$}}}}
\def\gsim{\mathrel{\raise.3ex\hbox{$>$\kern-.75em\lower1ex\hbox{$\sim$}}}}
\def\ifmath#1{\relax\ifmmode #1\else $#1$\fi}
\begin{document}
\begin{titlepage}
\begin{flushright}
{\tt hep-ph/0512169} \\
CLNS-05/1941 \\
FERMILAB-PUB-05-536-T \\
\end{flushright}

\vskip.5cm
\begin{center}
{\huge \bf The flavor of a little Higgs %\`a la 
with T-parity} \vskip.2cm
\end{center}

\begin{center}
{\bf {Jay Hubisz}$^a$, {Seung J. Lee}$^b$, {Gil Paz}$^b$} \\
\end{center}
\vskip 8pt
\begin{center}
$^{a}$ {\it Fermi National Accelerator Laboratory, \\ P.O. Box 500,
    Batavia, IL 60510} \\
\vspace{.3cm}
$^{b}$ {\it Institute for High Energy Phenomenology, Laboratory of Elementary Particle Physics, \\ 
Cornell University, Ithaca, NY 14853} \\ 
\vspace*{0.3cm} {\tt hubisz@fnal.gov, sjl18@cornell.edu, gilpaz@lepp.cornell.edu}

\end{center} 

\vglue 0.3truecm

\begin{abstract}
\vskip 3pt \noindent
We analyze flavor constraints in the littlest Higgs model with
T-parity.  In particular, we focus on neutral meson mixing in
the $K$, $B$, and $D$ systems due to one loop contributions from
T-parity odd
fermions and gauge bosons.  We calculate the short distance contributions to mixing for a
general choice of T-odd fermion Yukawa couplings.
We find that for a generic choice of textures,
a TeV scale GIM suppression is necessary to avoid large
contributions.  If order one mixing angles are allowed in the extended
flavor structure, the mass spectrum is severely constrained, and must
be degenerate at the $1$-$5$\% level.  However, there are still
regions of parameter space where only a loose degeneracy is necessary
to avoid constraints.  We also consider the $B_s$ system, and identify a
scenario in which the mixing can be significantly enhanced beyond the
standard model prediction while still satisfying bounds on the other
mixing observables.  We present both analytical and numerical
results as functions of the T-odd fermion mass eigenvalues.
\end{abstract}

\end{titlepage}

\newpage

\section{Introduction}
The mechanism of electroweak symmetry breaking (EWSB) will presumably be revealed in the coming years
through a combination of LHC and ILC data.  It is expected that
embedded in the newly discovered physics will be an explanation
of how this mechanism remains stable under quantum corrections.
Until this time, it is vital that we study the different known field
theoretical mechanisms of EWSB that stabilize the Higgs potential.

The little Higgs mechanism~\cite{oldLH,LHreview} is a revival of composite Higgs models~\cite{HiggsPseudo,KaplanGeorgi}
that attempted to solve these
issues.   In these models, the Higgs is a pseudo-Goldstone boson of
approximate global symmetries that are added on to the standard model (SM).  In the
little Higgs mechanism, the
electroweak scale is stabilized against quadratically divergent
corrections by the manner in which perturbative couplings break the
global symmetries.  In the simplest models, the Higgs mass receives no
quadratically divergent quantum corrections until two loop order,
although models with a larger symmetry structure can postpone these
corrections to higher loop order~\cite{oldLH}.

The most compact implementation of the little Higgs mechanism is known as
the littlest Higgs model~\cite{LH}.  In this model, the SM is enlarged to incorporate
an approximate $SU(5)$ global symmetry.  This symmetry is broken down
to $SO(5)$ spontaneously, though the mechanism of this
breaking is left unspecified.  The Higgs is an approximate Goldstone boson
of this breaking.

While the earliest littlest Higgs models have issues with low energy constraints~\cite{precision}, recent studies have shown that this structure is
still possible if one adds a discrete $Z_2$ symmetry to the model~\cite{LHT}.
Known as T-parity, this symmetry forbids the couplings which led to
stringent electroweak precision and compositeness bounds in the
original littlest Higgs model.

A consistent and phenomenologically viable littlest Higgs model with
T-parity requires the introduction of ``mirror fermions''~\cite{mirror}.  For each
new SM doublet, there must be another doublet which has the opposite
T-parity eigenvalue.  These mirror fermions are required to cut off
otherwise large four-fermion operators constrained primarily by LEP,
and Drell-Yan processes~\cite{kingman}, but they also open up a new
flavor structure in the model.  From studies of supersymmetry and other models of new physics,
it is known that new flavor structure at the TeV scale is quite
stringently constrained~\cite{SUSYflav}.  This is primarily due to the presence, in
the SM, of a GIM mechanism~\cite{SMGIM}.  The lightness of the SM fermions, coupled
with the near diagonal texture of the CKM matrix, strongly suppress flavor and
CP violating amplitudes, pushing them well below their naive dimension
analysis (NDA) estimated
values.  In the absence of a TeV scale GIM mechanism, new
contributions to neutral meson mixing and rare decays are often
many orders of magnitude larger than the SM contributions~\cite{MNGROSS,Nir}.

Neutral meson mixing, CP violation, and rare decays have been tested
experimentally through a variety of different observables, and are not
substantially different than expectations derived from SM calculations.
Therefore we expect there to be very little freedom in the new flavor
sector.  In this paper, we study the flavor constraints on the extended T-odd
fermion sector of the littlest Higgs model with T-parity.  Specifically, we
consider constraints from neutral meson particle anti-particle mixing,
leaving rare decays for future study.

In Section~\ref{model}, we outline the conventions used to derive the
Feynman rules relevant to flavor physics.  In Section~\ref{flavorsec},
we discuss how we approach the process of diagonalizing the action to
the mass eigenbasis, and identify the new parameters which
describe the new sources of flavor mixing and CP violation.  In
Section~\ref{smmix}, we outline the calculations for neutral meson
mixing in the SM.  In
Section~\ref{calculation}, the contributions to neutral meson mixing involving
the T-odd fields is presented.  Section~\ref{results} contains a numerical analysis of
the bounds on the parameters describing the T-odd fermion sector, and
an analysis of $B_s$ mixing.  We
conclude in Section~\ref{conclusions}.  In the Appendix, we give the
relevant Feynman rules, as well as the formulas which arise from calculating
the one loop contributions to flavor changing operators.

\section{The Model}
\label{model}

The littlest Higgs model~\cite{LH} is the most compact way of extending the
SM to include a collective symmetry breaking structure that protects
the Higgs mass.  In the
littlest Higgs model, the theory is approximately invariant under
$SU(5)$ global symmetry transformations.  A scalar VEV of an $SU(5)$
symmetric tensor $\Sigma$ spontaneously
breaks this $SU(5)$ down to $SO(5)$ at a scale $f$.  This scale is
presumed to be $\mathcal{O}(1\ \mathrm{TeV})$.  The Higgs boson is
one of the Goldstone bosons associated with this
breaking.  An $\left[ SU(2) \times U(1)
  \right]^2$ subgroup embedded in the global $SU(5)$ is gauged, and after
$\Sigma$ gets a VEV, this gauge symmetry is reduced to the SM $SU(2)_L
\times U(1)_Y$.  Perturbative couplings in the model break the $SU(5)$
global symmetry explicitly, and quantum corrections involving these
interactions generate masses and non-derivative couplings for the
Goldstone fluctuations, rendering them pseudo-Goldstone bosons.  

The Higgs mass is protected from quadratic divergences at the one-loop
level due to the way in which perturbative couplings are introduced.  
Any single coupling preserves at least one of two overlapping $SU(3)$
subgroups of the full $SU(5)$ global symmetry.  Under these $SU(3)$
subgroups, the Higgs is still an exact Goldstone boson.  The VEV which
breaks the $SU(5)$ softly breaks these $SU(3)$ symmetries, and thus
generates logarithmically divergent contributions to the Higgs mass
at one loop.  Amplitudes involving perturbative couplings only
generate a quadratically divergent contribution at two loop order.
The value for the Higgs mass obtained by NDA arguments is then
suppressed relative to the breaking scale $f$ by a loop factor.

The effective action is parametrized by a non-linear sigma model.
Only the Goldstone bosons of the $SU(5)$ breaking
are included in the low energy effective theory, and the way in which
the theory is linearized, or UV completed, is left
ambiguous.  In terms of these Goldstone fields, the symmetric tensor
$\Sigma$ can be expressed as:
\begin{equation}
\Sigma = e^{2 i \Pi/f} \Sigma_0.
\end{equation}
The ``pion'' matrix $\Pi$ contains the Goldstone degrees of freedom, and
$\Sigma_0$ is the VEV of $\Sigma$:
\begin{equation}
\Sigma_0 = \left(\begin{array}{ccccc}
0 & 0 & 0 & 1 & 0 \arline
0 & 0 & 0 & 0 & 1 \arline
0 & 0 & 1 & 0 & 0 \arline
1 & 0 & 0 & 0 & 0 \arline
0 & 1 & 0 & 0 & 0 \end{array}\right).
\end{equation}

To implement the collective symmetry breaking structure, the gauge
generators are embedded in the $SU(5)$ global symmetry such that
any given generator commutes with an $SU(3)$ subgroup of the
$SU(5)$ global symmetry:
\begin{eqnarray}\label{gauged}
&Q_1^a=\left( \begin{array}{ccc} \sigma^a/2 &0 & 0 \\
0 & 0 & 0\\ 0 & 0 & 0
\end{array}\right), \ \ \ &Y_1=
{\rm diag}(3,3,-2,-2,-2)/10\nonumber \\
&Q_2^a=\left( \begin{array}{ccc} 0 & 0 & 0\\
0 & 0 & 0 \\
0 &0&-\sigma^{a*}/2\end{array} \right), & Y_2={\rm
diag}(2,2,2,-3,-3)/10~.
\end{eqnarray}
The $Q_1$ and $Y_1$ generators commute with the $SU(3)_2$ subgroup of
$SU(5)$ whose generators occupy the lower right hand corner.  The
$Q_2$ and $Y_2$ generators similarly commute with the $SU(3)_1$
subgroup in the upper left.

The VEV $\Sigma_0$ breaks the extended gauge group $\left[ SU(2)
  \times U(1) \right]^2$ down to the SM electroweak
  $SU(2)_L \times U(1)_Y$, leading to the broken combinations
acquiring masses given to lowest order in $v/f$ by
\begin{equation}
M_{W_H} = g f, \ M_{Z_H} = g f,\ M_{A_H} = \frac{g' f}{\sqrt{5}}.
\end{equation}
The pseudo-Goldstone bosons of the $SU(5)$
  breaking then decompose
  into representations of the electroweak gauge group as follows:
\begin{equation}
\mathbf{1_0} \oplus \mathbf{3_0} \oplus \mathbf{2_{1/2}}
\oplus \mathbf{3_{1}}.
\end{equation}
The $\mathbf{1_0}$ and $\mathbf{3_0}$ are eaten in the Higgsing of
the extended gauge sector down to the SM gauge group.

The pion matrix, with the Higgs doublet and complex triplet $\phi$
identified along with the eaten Goldstone bosons, is given by
\begin{equation}
\Pi\,= \left(\begin{array}{ccccc}
-\omega_3/2- \eta/\sqrt{20} & -\omega^+/\sqrt{2} & -i \pi^+/\sqrt{2} & -i 
\phi^{++} & -i
\frac{\phi^{+}}{\sqrt{2}} \\
-\omega^-/\sqrt{2} & \omega_3/2- \eta/\sqrt{20} & \frac{v+h+i \pi^0}{2}& -i
\frac{\phi^{+}}{\sqrt{2}} & \frac{-i \phi^0 +\phi_P^0}{\sqrt{2}} \\
i \pi^-/\sqrt{2} & (v+h-i \pi^0)/2 & \sqrt{4/5} \eta  & -i
\pi^+/\sqrt{2} & (v+h+i \pi^0)/2 \\
i \phi^{--} & i \frac{\phi^{-}}{\sqrt{2}} & i \pi^-/\sqrt{2} &
-\omega_3/2 - \eta/\sqrt{20} &
- \omega^-/\sqrt{2} \\
i \frac{\phi^{-}}{\sqrt{2}} & \frac{i \phi^0 +\phi_P^0}{\sqrt{2}} &
\frac{v+h-i \pi^0}{2} & - \omega^+/\sqrt{2} &\omega_3/2- \eta/\sqrt{20}
\end{array}\right).
\label{pions}
\end{equation}

In the model we consider, a T-parity $Z_2$ discrete symmetry is enforced to make the model
consistent with electroweak precision tests.  This $Z_2$ is derived
from an
automorphism of the gauge groups which exchanges the $\left[SU(2)
\times U(1)
  \right]_1$ gauge group with $\left[SU(2) \times U(1)
  \right]_2$.  If the Lagrangian is made invariant under such a
transformation, tree
level electroweak precision constraints are
avoided~\cite{LHT,LHTprec}.  This can be achieved by setting couplings
associated with the two gauge groups to be equal, and also imposing
that the particle content of the model is symmetric under this
transformation.  If the
symmetry is made exact, the lightest T-parity odd particle is
stabilized, and is a dark matter candidate~\cite{LHT,LHTpheno}.  The
heavy gauge bosons are odd under T-parity, and so tree level four-fermion
operators involving SM fermions are also forbidden.

Under T-parity, the Goldstone boson matrix transforms as
\begin{equation}
T: \Pi \rightarrow - \Omega \Pi \Omega
\end{equation}
where $\Omega = \mathrm{diag}(1,1,-1,1,1)$.  This transformation law
can be derived from the requirement that the kinetic term for $\Sigma$
be invariant under exchange of the two sets of gauge bosons.  This transformation
law for the Goldstone bosons ensures that the $SU(2)_L$
triplet is odd under
T-parity, and that there is thus no trilinear coupling of the
triplet to the SM Higgs doublet.  This forbids a small VEV being
generated for the triplet which would otherwise cause
phenomenologically constrained violations of the custodial $SU(2)$
symmetry of the SM Higgs potential~\cite{precision}.

\subsection{Fermion Content}
We will give now in detail the structure of the
fermion sector of the model.  To avoid compositeness constraints and
simultaneously implement T-parity, it is necessary to double the SM fermion
doublet spectrum~\cite{mirror}.  For each SM $SU(2)_L$ doublet, a doublet
under $SU(2)_1$ and one under $SU(2)_2$ are introduced.  The T-parity even
combination is associated with the SM $SU(2)_L$ doublet while
the T-odd combination is given a mass of order the breaking scale,
$f$. The fermion doublets $\psi_1,\psi_2$ can be embedded into
incomplete representations $\Psi_1,\Psi_2$ of $SU(5)$, and the
field content can be expressed as follows:
\begin{equation}
\begin{array}{ccc}
\Psi_1=\left(\begin{array}{c} \psi_1 \\ 0 \\ 0 \end{array}\right),
& \Psi_2=\left(\begin{array}{c} 0 \\ 0 \\ \psi_2
\end{array}\right), &
\tilde{\Psi} = \left(\begin{array}{c} \tilde{\psi}_R \\ \chi_R \\ \psi_R
\end{array}\right),
\end{array}
\end{equation}
where $\tilde{\Psi}$ is a T-odd $SO(5)$ multiplet which transforms
non-linearly under the global $SU(5)$.  The transformation laws for $\Psi_1$ and $\Psi_2$
under $SU(5)$ are as follows:
\begin{equation}
\Psi_1 \rightarrow V^* \Psi_1 \hspace{.2in} \Psi_2 \rightarrow V
\Psi_2,
\end{equation}
where $V$ is an $SU(5)$ transformation.  The action of T-parity on the multiplets takes $\Psi_1\to
-\Sigma_0 \Psi_2$ and $\tilde{\Psi} \to -\tilde{\Psi}$.  It is
possible to extend the gauge and global symmetry structure of the
model to include new T-even gauge bosons and scalars, and in some of
these extensions, all of the fermions that are
introduced can be made to transform linearly~\cite{mirror}.  The
flavor changing processes that we calculate are in fact present in all of
these models.  We note however that in these extensions, there may be new
flavor changing processes involving the extra T-even fields that give
additional contributions.  We choose to work with the
model that has the simplest gauge and global symmetry structure, and
which is likely the least constrained.

The T-parity even combination of $\psi_1$ and $\psi_2$ are the
SM electroweak quark and lepton doublets, while the T-odd
combination is given a Dirac mass with the $\psi_R$ of the
$\tilde{\Psi}$ $SO(5)$
representations through the following Yukawa interaction:
\begin{equation}\label{yukdiag}
\kappa f \left( \bar{\Psi}_{2} \xi \tilde{\Psi} +
\bar{\Psi}_{1} \Sigma_0 
\Omega \xi^\dagger \Omega \tilde{\Psi} \right)+ {\mathrm h.c.}
\end{equation}
The insertion of $\xi = e^{i \Pi/f}$ is necessary to make these terms
invariant under $SU(5)$ rotations~\cite{LHT,CCWZ}.  The T-odd combination of
left-handed doublets gains a mass (before EWSB) equal to $\sqrt{2} \kappa f$.  After
EWSB, a small mass splitting between the T-odd up
and down-type quarks is induced, and the masses are given by
\begin{eqnarray}
&&m_{d_-} = \sqrt{2} \kappa f \arline
&&m_{u_-} = \sqrt{2} \kappa f \left(1- \frac{1}{8}
\left(\frac{v}{f}\right)^2 + \cdots \right).
\end{eqnarray}

The remaining degrees of freedom
in $\tilde{\Psi}$ are given masses with another non-linearly
transforming multiplet.  A spinor multiplet of $SO(5)$
could be introduced.  This multiplet includes two $SU(2)_L$ singlets,
and one doublet.  These could marry with $\chi_R$ and
$\tilde{\psi}_R$, and then one more singlet would be necessary to lift
the entire spectrum.

We note that the $\tilde{\Psi}$ fields do not have any
gauge interactions with SM fermions, due to the fact that they are odd
T-parity eigenstates.  They can thus only have gauge interactions with
the T-even SM gauge bosons.  Similarly, the interactions with T-odd
Goldstone bosons coming from Eq.~(\ref{yukdiag}) can only involve the
left-handed SM fermions.

In our analysis, we assume that the mass of the doublets
$\tilde{\psi}_R$ are much larger than the breaking scale $f$, and the
additional singlets $\chi_R$ have masses of $\sim 5 f$.  They are necessary to cancel two loop
quartic divergences to the Higgs mass, but are otherwise allowed to
be decoupled from the spectrum~\cite{LHTpheno}.  At $5 f$, the mass is
large enough to have only negligible effects on low energy
phenomenology, but low enough to keep the Higgs mass small.  The $\tilde{\psi}_R$
doublet is necessary only to cancel a divergence proportional to
$g'^2$, which is relatively small, so it is fine for its mass to be
rather large (perhaps $10$~TeV).  Increasing its mass also decouples its
effects on flavor physics, as the masses are not due to the Yukawa
coupling $\kappa$, and are simple Dirac masses.  The flavor changing couplings of the
$\chi_R$ singlets arise only at order $v/f$, and thus the effects are
suppressed relative to those we calculate.  In summary, including these fermions in the flavor analysis
is a higher order effect.  We note that if the $\tilde{\psi}_R$
doublet is taken to be light, then its flavor effects arise through
box diagrams where components of the complex triplet $\phi$ run in the loop.

In order to prevent against large contributions to the Higgs mass
from one loop quadratic divergences, the third generation light Yukawa
interaction must be modified so that it incorporates the collective
symmetry breaking structure.  In order to do this, the $\Psi_1$ and
$\Psi_2$ multiplets for the third generation must be completed to
representations of the $SU(3)_1$ and $SU(3)_2$ subgroups of
$SU(5)$.  These multiplets are
\begin{equation}
\begin{array}{ccc}
Q_1=\left(\begin{array}{c} q_1 \\ t'_1 \\ 0 \end{array}\right), &
Q_2=\left(\begin{array}{c} 0 \\ t'_2 \\ q_2
\end{array}\right),
\end{array}
\end{equation}
where $Q_1$ and $Q_2$ obey the same transformation laws under
T-parity and the $SU(5)$ symmetry as do $\Psi_1$ and $\Psi_2$.  It
should be noted that the quark doublets are embedded such that
\begin{equation}
q_i = -i \sigma_2 \left(\begin{array}{c} t_i \\ b_i
\end{array}\right).
\end{equation}
One must also introduce additional singlets $t'_{1R}$ and
$t'_{2R}$ which transform under T-parity as
\begin{equation}
t'_{1R}\rightarrow -t'_{2R}
\end{equation}
so the top sector can be implemented in the following T-parity
invariant way~\cite{LHT,mirror}
\begin{eqnarray}\label{topyuk}
\mathcal{L}_t &=& \frac{1}{4}\lambda_1 f \epsilon_{ijk}
\epsilon_{xy} \big[ (\bar{Q}_1)_i \Sigma_{jx} \Sigma_{ky}  -
(\bar{Q}_2 \Sigma_0)_i \tilde{\Sigma}_{jx} \tilde{\Sigma}_{ky}
\big] u_{3R} \arline && \hspace{1in} + \lambda_2 f (\bar{t}'_1
t'_{1R} + \bar{t}'_2 t'_{2R})+ h.c.
\end{eqnarray}
  This Yukawa interaction generates a mass for the top quark given by
\begin{equation}
m_{\mathrm{top}} =\frac{ \lambda_1 \lambda_2
v}{\sqrt{\lambda_1^2+\lambda_2^2}},
\end{equation}
while the orthogonal T-even combination ($T_+$), and the T-odd combination of
$t'_1$ and $t'_2$, ($T_-$) acquire masses given by
\begin{equation}
m_{T_+} = \sqrt{\lambda_1^2+\lambda_2^2} f,\ \ \mathrm{and}\ \ m_{T_-}
= \lambda_2 f.
\end{equation}
The T-odd
combination of the $q_1$ and $q_2$ doublets get their mass from the
same Yukawa coupling as the other T-odd doublets discussed earlier.  The other
two generations of SM up-type quarks acquire their
mass through similar terms, though with the $t'$ quarks missing
from the $Q_1$ and $Q_2$ multiplets since the Yukawa couplings are
small and these quadratic divergences are suppressed.  The $T_-$ quark
only has sizeable ``flavor changing'' interactions with the SM top quark
mass eigenstate and the $A_H$~\cite{LHTpheno}, and so it does not
contribute to any of the processes we study.

\section{T-odd Flavor Mixing}
\label{flavorsec}
Before beginning a discussion of the T-odd fermion mass sector, we
briefly review the process as it works in the SM~\cite{CKM}.  The Yukawa sector
generates mass matrices for the three up-type quarks given by
$M_{uj}^{i}$ after EWSB which is diagonalized by two unitary matrices,
$U$ and $V$:
\begin{equation}
(M_{u})^i_j = (V_u)^i_k (M_{u}^D)^k_l (U_u^\dagger)^l_j
\end{equation}
The gauge eigenstates are then expressed in terms of (the primed) mass
eigenstates by 
\begin{equation}
u_L^i = (V_u)^i_j u'^{j}_L\ \ \ u_R^i = (U_u)^i_j u'^{j}_R.
\end{equation}
A similar procedure applies to the down-type quark mass matrix.  Much
of the information contained in the diagonalization of the mass matrices is
redundant when one looks at SM amplitudes for cross
sections.  The cross-over to the mass eigenbasis leaves most of the gauge
interaction portion of the Lagrangian invariant.  It is only the weak
interactions which couple the $T_3=1/2$ and $T_3 = -1/2$ sectors that are affected:
\begin{equation}
\frac{g}{\sqrt{2}} \left[ \bar{u}^i \dslash{W}^+ P_L d^i + \bar{d}^i
  \dslash{W}^- P_L u^i \right] = \frac{g}{\sqrt{2}} \left[ \bar{u'}_i
  (V_u^{\dagger})^i_j \dslash{W}^+ P_L (V_d)^j_k d'^k + \bar{d}_i
  (V_d^\dagger)^i_j 
  \dslash{W}^- P_L (V_u)^j_k u'^k \right]
\end{equation}
In the SM, the only observable rotation is the combination
\begin{equation}
(V_u^\dagger)^i_k (V_d)^k_j \equiv (V_{\mathrm CKM})^i_j.
\end{equation}
This is no longer necessarily the case when one introduces additional
fermions which couple to the SM.

The mass eigenbasis in the T-odd fermion sector is not necessarily
aligned with the SM fermion sector.  These additional mixings are
a source of flavor changing processes that are the focus of this
paper.  The interaction that gives the T-odd doublets their mass,
Eq.~(\ref{yukdiag}), can be extended to include generational mixing:
\begin{equation}\label{toddyukmix}
\kappa^i_j f \left( \bar{\Psi}_{2i} \xi \tilde{\Psi}^j +
\bar{\Psi}_{1i} \Sigma_0 
\Omega \xi^\dagger \Omega \tilde{\Psi}^j \right)+ {\mathrm h.c.}
\end{equation}
In analogy with the CKM transformations, the resulting mass matrix $\sqrt{2} f
\kappa^i_j$ is diagonalized by two $U(3)$ matrices:
\begin{equation}
\kappa^i_j = (V_H)^i_k (\kappa_{D})^k_l (U_H^\dagger)^l_j.
\end{equation}
$V_H$ acts on the left handed fields while $U_H$ acts on the right
handed $\tilde{\Psi}$ fields.  We note that these matrices are
identical for the up and down-type T-odd fermions, since the resulting
Dirac mass terms are $SU(2)_L$ symmetric.

The gauge interaction portion of the kinetic terms in the T-parity
eigenbasis are given qualitatively by 
\begin{equation}
g \bar{Q}_{-i} \dslash{A_-} Q_+^i + g \bar{Q}_{+i} \dslash{A_-} Q_-^i,
\end{equation}
where the $A_-$ and $Q_-$ are the T-odd gauge bosons and fermions with
mass $\sim f$.  The $Q_+$ are the T-even eigenstates.
One can further rotate this T-parity eigenbasis into the mass
eigenbasis, where flavor mixings in both the T-odd and T-even sectors
are taken into account.  Identifying the mass eigenstates with a $H$
and $L$ index for heavy and light, respectively, these interactions
can be re-expressed as
\begin{equation}
g \bar{Q}_{Hi} V_{Hj}^{\dagger i} \dslash{A_H} \left(\begin{array}{c}
  (V_u)^j_k u_L^k \\ (V_d)^j_k d_L^k \end{array}\right) + g \left(\begin{array}{c}
   \bar{u}_{Lk} (V_u^\dagger)^k_i \\  \bar{d}_{Lk} (V_d^\dagger)^k_i \end{array}\right)
\dslash{A_H} V_{Hj}^{i} Q_H^j,
\end{equation}
where
\begin{equation}
Q_H^i = \left(\begin{array}{c}
u_H^i \arline
d_H^i
\end{array}\right).
\end{equation}
The rotation matrix $V_H$ is in $U(3)$, and operates on the flavor
indices of the left handed T-odd fermions.  In analogy with the CKM
matrix then, the rotations relevant to flavor physics are
\begin{equation}
(V^\dagger_H)^i_k (V_u)^k_j \equiv (V_{Hu})^i_j, \ \ (V^\dagger_H)^i_k
  (V_d)^k_j \equiv (V_{Hd})^i_j.
\end{equation}
Note that the two matrices are related through the SM CKM matrix:
\begin{equation}
V_{Hu}^\dagger V_{Hd} = V_{\mathrm{CKM}}.
\end{equation}
This is an important result, as it implies that one cannot completely
turn off the new mixing effects except with a universally degenerate mass
spectrum for the T-odd doublets.  For example, if $V_{Hd}$ is set to be the
identity, then $V_{Hu}^\dagger = V_{{\rm CKM}}$.

There is a subtlety here involving the T-even partner of the top quark
which is responsible for canceling the top quark's quadratically
divergent contribution to the Higgs mass.  As it is only inserted in
the top quark sector, it explicitly breaks flavor symmetries in a way
such that the symmetry cannot be restored through a spurion analysis.
If we were to assume that the
up-type Yukawa couplings are flavor diagonal, then the top quark
divergence is canceled as in the littlest Higgs model.  From this
starting point, where $V_{u} = {\bf 1}$, $V_{Hu} = V_H^\dagger$, and $V_{Hd} =
V_H^\dagger V_d$, and $V_d = V_{{\rm CKM}}$.  Because there is no
symmetry in place which forbids such off-diagonal top-Yukawa elements, this
perhaps seems a bit unnatural.

We note that the flavor symmetry could easily be restored by
completing all three generations of the $SU(2)_1$ and $SU(2)_2$
doublet quarks to be $SU(3)_1$ and $SU(3)_2$ triplets.  Doing so leads
to a somewhat more natural picture of how the top quark divergence is
canceled, but at the expense of introducing $4$ additional particles
(a T-even and T-odd partner for each of the two remaining up-type
quarks).  The effects of these new quarks on flavor physics and EWP would be
vanishingly small, since, as found in~\cite{LHTprec,buras}, these effects are
approximately proportional to $m^{f\ 4}_{SM}/m^4_{T_+}$, where
$m_{SM}^f$ is the mass of either the up or charm quark.  The new $T_-$
flavor contributions would scale in the same way.  Since the masses
for the first two generations are quite small,
this effect is extremely suppressed.  Depending on the mass of these new
particles, however, the collider phenomenology~\cite{LHTpheno,maximcollider,Han,maximtopquarks,Atlas} could be quite
different.  For the remainder of this analysis, we assume that the
flavor symmetry is only explicitly violated by mass terms, and that
the fermions come in $SU(3)_i$ multiplets, and that therefore, $V_u$
is free to take on any value.  The earlier model can
easily be obtained from this one by picking specific mass textures,
and decoupling the partners of the lighter up-type quarks.

Beyond the SM, there are three new rotation angles, and
one new CP violating phase, as we explain here.  There are two unitary matrices which show
up in observables, $V_{Hu}$, and $V_{Hd}$.  These have 3 rotations each, and
6 phases each.  There are 6 quark fields which transform under
$SU(2)_1$, and 6 under $SU(2)_2$.  Each set of 6 quark fields can
absorb 5 phases (an overall phase in each sector is
unobservable).  What remains are 6 total rotations, and 2 CP violating
phases.  One combination of $V_{Hu}$ and $V_{Hd}$ gives the SM CKM
matrix, which has 3 rotations and 1 phase.  We then parametrize $V_{Hd}$ the same
way as we do the CKM matrix, but with new angles $\theta_{12}^{d}$, $\theta_{23}^{d}$, $\theta_{13}^{d}$, and phase $\delta_{13}^{d}$:
\begin{equation}
V_{Hd} = \left( \begin{array}{ccc}
c^{d  }_{12} c^{d  }_{13} & s^{d  }_{12} c^{d  }_{13} & s^{d  }_{13} e^{-i \delta^{d  }_{13}} \\
-s^{d  }_{12} c^{d  }_{23}-c^{d  }_{12} s^{d  }_{23} s^{d  }_{13} e^{i \delta^{d  }_{13}} & c^{d  }_{12} c^{d  }_{23} - s^{d  }_{12} s^{d  }_{23} s^{d  }_{13} e^{i \delta^{d  }_{13}} & s^{d  }_{23} c^{d  }_{13} \\
s^{d  }_{12} s^{d  }_{23} - c^{d  }_{12} c^{d  }_{23} s^{d  }_{13} e^{i \delta^{d  }_{13}} & - c^{d  }_{12} s^{d  }_{23} - s^{d  }_{12} c^{d  }_{23} s^{d  }_{13} e^{i \delta^{d}_{13}} & c^{d}_{23} c^{d}_{13}
\end{array} \right).
\label{newCKM}
\end{equation}
The matrix $V_{Hu}$ can then be extracted from the
relation $V_{Hu} = V_{Hd} V^\dagger_{{\rm CKM}}$.  With this
parametrization, we can
analyze all of the physical degrees of freedom in the model.
Throughout our analysis, we use for the SM CKM
matrix the PDG best fit angles~\cite{PDG}
\begin{equation}
s_{12} = 0.2243\pm0.0016,\ s_{23} = 0.0413\pm0.0015,\ s_{13}= 0.0037\pm0.0005,\
\delta_{13} = 1.05\pm0.25. 
\end{equation}

There are also interaction terms containing a T-odd Goldstone boson, a
T-odd fermion, and a SM fermion.  These arise
from expanding the T-odd Yukawa interactions in Eq.~(\ref{toddyukmix})
in the mass eigenbasis.  Similarly, these only
involve the rotations $V_{Hu}$ and $V_{Hd}$.  In
Table~\ref{flavorrules!}, in the Appendix, we give the Feynman rules
relevant to flavor physics. 

\section{Mixing in the Standard Model and Beyond}
\label{smmix}
The state of the art theory predictions for mixing in the $K$ and $B_d$ systems
agree with experimental results up to theoretical errors in long
distance effects and QCD
corrections.  In $D$ mixing, there is only an upper bound.  In this section, we give a very brief summary of
the SM predictions for neutral meson mixing.  For more detailed discussions
see~\cite{weakdecrev,Buras:2001pn,Battrev,BurasCPVIO,Buras:2005xt}.
We also comment on the relevance of each system to our study of the
littlest Higgs model with T-parity.

\subsection{Standard model effective Hamiltonian}

The lowest order SM contribution to the effective
Hamiltonian that governs neutral $K$ meson mixing is given by
\cite{Inami:1980fz}
\begin{equation}
\mathcal{H}^\mathrm{SM}_\mathrm{eff} = \frac{G_F^2}{16 \pi^2}
M_{W_L}^2 \sum_{ij}\lambda_i \lambda_j F(x_i,x_j\,;M_{W_L})(\bar{s}
d)_{(V-A)} (\bar{s} d)_{(V-A)},
\end{equation}
where $x_i=m_i^2/M_{W_L}^2$, $m_i$ and $m_j$ are the masses of the
quarks in the loop, and the $\lambda_i$
are defined as functions of CKM matrix elements:  $\lambda_i=V_{\rm
  CKM}^{\,is}V_{\rm CKM}^{\,*id}$.  
The function $F(x_i,x_j;M_{W_L})$ is given in the
't Hooft-Feynman gauge in the Appendix.  This function is finite in the 't
Hooft-Feynman gauge, but divergent in unitary gauge. When summing
over the different flavors, the gauge dependence cancels after imposing
unitarity of the CKM matrix through the relation $\lambda_u = -
\lambda_c - \lambda_t$.  The final form of the effective
Hamiltonian after this substitution is given by
\begin{eqnarray}
\mathcal{H}^\mathrm{SM}_\mathrm{eff} &=& \frac{G_F^2}{16 \pi^2}
M_{W_L}^2
\left[ \lambda^2_c
  \,\eta_1\, \widetilde{S}_0(x_u,x_c)+2\,
  \lambda_c \lambda_t \eta_3\, \widetilde{S}_0(x_u,x_c,x_t) + \lambda^2_t\,
  \eta_2\, \tilde{S}_0 (x_u,x_t) \right] \arline
&& \hspace{3in}\times (\bar{s} d)_{(V-A)} (\bar{s} d)_{(V-A)},
\label{effhamSM}
\end{eqnarray}
where
\begin{eqnarray}
&& \widetilde{S}_0 (x_i, x_j)= F (x_i,x_i\,;M_{W_L}) - 2 F(x_i,
x_j\,;M_{W_L}) + F(x_j,x_j\,;M_{W_L}) \arline
&& \widetilde{S}_0 (x_i, x_j,
x_k)= F (x_i,x_i\,;M_{W_L}) - F(x_i,x_j\,;M_{W_L}) -
F(x_i,x_k\,;M_{W_L})+ F(x_j,x_k\,;M_{W_L}), \arline
&&\ 
\end{eqnarray}
and the $\eta_i$ are QCD corrections. 
%Our notation is such that the index
%of $\lambda_i$ refers to the appropriate quark in the $\mathrm{i^{th}}$
%generation.
We will see a similar structure for 
the contributions to the effective Hamiltonian from the T-odd fermions.  In practice, for the SM particles the masses
of the lighter particles are taken to be zero in the formula above,
leading to simplified expressions. For example, in $K$ and $B$ mixing, 
taking $m_u=0$ gives the standard functions:
\begin{equation}
S_0(x_c,x_t)=\widetilde{S}_0(0,x_c,x_t),\quad S_0(x_t)=\widetilde{S}_0(0,x_t).
\end{equation}

The effective Hamiltonians for mixing in the other neutral meson
systems can easily be obtained from Eq.~(\ref{effhamSM}) by altering
the $\lambda_i$, the four-quark operator, and the $\eta_i$.  For example,
to get the result for $B_d$ mixing, each occurrence of $s$ should be
replaced by $b$, and the $\eta$'s that correspond to the $B$
system should be inserted.

\subsection{Physical Observables}
We show here how to obtain the physical observables from the effective
Hamiltonians.  We restrict our analysis to neutral meson mass
splittings and $\epsilon_K$.  We comment on the relevance of these
observables to our analysis of T-parity flavor physics.

\subsubsection{$K^0-\bar{K}^0$ Mixing}
We will use two observables from $K^0-\bar{K}^0$ Mixing: the mass
difference $\Delta M_K$, and the parameter $\epsilon_K$, related to
the real and the imaginary part of $\langle
K^0|\mathcal{H}_\mathrm{eff}|\bar{K}^0\rangle$,
respectively. More specifically we have for $\Delta M_K$:
\begin{equation}
\Delta M_{K}=\frac{1}{m_K}\mathrm{Re}\,\langle
K^0|\mathcal{H}_\mathrm{eff}|\bar{K}^0\rangle.
\end{equation}
The SM prediction is
\begin{equation}
\Delta M_{K}=\mathrm{Re}\left\{\frac{G_{\rm F}^2}{6 \pi^2} F_K^2 \hat
B_K m_K M_{W_L}^2 \left[ {\lambda_c^*}^2 \eta_1 S_0(x_c) +
{\lambda_t^*}^2 \eta_2 S_0(x_t) + 2 {\lambda_c^*} {\lambda_t^*} \eta_3
S_0(x_c, x_t) \right]\right\},
\end{equation}
where $F_K$ and $m_K$ are the $K$-meson decay constant and mass,
respectively. $\hat B_K$ is an order one non-perturbative ``bag'' parameter.
%, defined as
%($J_3=1.895$ in the NDR scheme):
%\begin{equation}
%\hat B_K = B_K(\mu) \left[ \alpha_s^{(3)}(\mu) \right]^{-2/9} \,
%\left[ 1 + \frac{\alpha_s^{(3)}(\mu)}{4\pi} J_3 \right]~,
%\end{equation}
%\begin{equation}
%\langle \bar K^0| (\bar s d)_{V-A} (\bar s d)_{V-A} |K^0\rangle
%\equiv \frac{8}{3} B_K(\mu) F_K^2 m_K^2
%\end{equation}

%The experimentally measured value of the mass difference is
%\begin{equation}
%\Delta M_K=M(K_{\rm L})-M(K_{\rm S}) = 
%(3.483\pm 0.006) \cdot 10^{-15} {\rm GeV}\,.
%\end{equation}

The theoretical prediction for $\epsilon_K$ is given by 
\begin{equation}
\epsilon_K\approx \frac{\exp(i \pi/4)}{2 \sqrt{2}\, \Delta M_K\, m_K} \,
 \mathrm{Im}\,\langle
K^0|\mathcal{H}_\mathrm{eff}|\bar{K}^0\rangle,
\end{equation}
and the SM prediction is then
\begin{equation}
\epsilon_k=\frac{G_{\rm F}^2 F_K^2 m_K M_{W_L}^2}{6 \sqrt{2} \pi^2\,
 \Delta M_K} \hat{B}_K \mathrm{Im}\,\lambda_t \left\{
 \mathrm{Re}\,\lambda_c \left[ \eta_1 S_0(x_c) - \eta_3 S_0(x_c, x_t)
 \right] - \mathrm{Re}\,\lambda_t \eta_2 S_0(x_t) \right\} \exp(i
 \pi/4)\,.
\end{equation}

$K$ mixing imposes some of the tightest bounds on the T-odd
fermion spectrum.  Bounds on the mass splitting of the neutral mass
eigenstates (or, equivalently, the mixing frequency of the CP
eigenstates) impose constraints on the first two generations of T-odd
fermions, as we will show in Section~\ref{results}.  In addition, if
there is a CP violating phase in $V_{Hd}$, then there are new physics
contributions to the $\epsilon_K$ observable, the measure of indirect
CP violation in $K$ decays.  As we show in
Section~\ref{results}, this observable is often the
most sensitive to little Higgs physics.  This is in analogy with the
``$\epsilon_K$ problem'' in supersymmetry (see for
example~\cite{Nir}).  In addition, it would be interesting to study
the $\epsilon'$ observable, which is the measure of direct CP
violation in $K$ decays.  We leave this for future work.

%The experimental value of $\epsilon_K$ is \cite{Eidelman:2004wy}:
%\begin{equation}
%\epsilon_{K}
%=(2.284\pm0.014)\cdot10^{-3}\;\exp{i\pi\over 4},
%\end{equation}

\subsubsection{$B_q^0-\bar{B}_q^0$ Mixing}

In our analysis we will also discuss the mass differences in the
$B^0_d-\bar{B}^0_d$ and $B^0_s-\bar{B}^0_s$ systems.  In terms of low
energy matrix elements, these mass splittings are given by
\begin{equation}
\Delta M_{B_q}=\frac{1}{m_{B_q}}\mathrm{Re}\,\langle
B_q^0|\mathcal{H}_\mathrm{eff}|\bar{B}_q^0\rangle.
\end{equation}
For neutral B mesons the
functional form is identical to the effective Hamiltonian for $K$
mixing, although with new $\lambda_i$, $\eta_i$, and bag parameters.
The hierarchy of the CKM matrix elements, however, allows
for simplification of the effective Hamiltonian, so that to an
excellent approximation, it depends only on $S_0(x_t)$.
The SM prediction is therefore
\begin{equation}
\Delta M_q = \frac{G_{\rm F}^2}{6 \pi^2} \eta_B m_{B_q} 
(\hat B_{B_q} F_{B_q}^2 ) M_{W_L}^2 S_0(x_t) \mathrm{Re}(\lambda^{*2}_t).
\end{equation}

Neutral $B$ mixing is particularly interesting due to the large amount of
progress currently being made both on experimentally constraining
$b$-quark physics, and on pinning down the theoretical SM predictions
for $B$ meson observables.  A particularly exciting system to study
from the perspective of current developments is the $B_s$ system.  The
mass splitting is so far undetermined by experiment.

In our analysis, the $B_d$ neutral meson mass splitting
provides constraints that are complementary to those from the $K$
system.  Because the $B_d$ system is more sensitive to physics in the
third generation, it generally imposes stronger bounds on the third
generation T-odd fermion doublet than the $K$ system alone.

%The experimentally measured values of these mass differences are 
%\begin{eqnarray}
%\Delta M_{B_d}&=&M(B^0_{d\,H})-M(B^0_{d\,L}) = (3.3044\pm 0.046) \cdot
%10^{-13} {\rm GeV}\nonumber\\
%\Delta M_{B_s}&=&M(B^0_{s\,H})-M(B^0_{s\,L}) > 94.8 \cdot
%10^{-13} {\rm GeV}, {\rm CL}= 95\%
%\end{eqnarray}

\subsubsection{$D^0-\bar{D}^0$ Mixing}
For $D$ meson mixing no mass splitting has yet been observed.  The SM
short distance 
contribution to the $D$ mixing effective Hamiltonian is extremely
suppressed, due to GIM and CKM factors.  There are potentially larger
long distance contributions, but these are not well understood due to
sensitivity to low energy strong dynamics.  The current
experimental bound is given by~\cite{PDG}
\begin{equation}
|m_{D_1^0}-m_{D_2^0}|< 4.6 \cdot 10^{-14}\ {\rm GeV},\ \  {\rm CL}= 95\%.
\end{equation}
In our analysis, we assume that the new physics contribution
dwarfs any SM contributions.

In our analysis, the $D$ system provides an important counterweight in
constraining the extended fermion sector.  Unlike the $K$ and $B$
systems, the T-odd fermion contributions to
mixing come from the up-type diagonalization matrix, $V_{Hu}$.  Without
the current experimental upper bound on the $D$ meson mass splitting, the
constraints on the T-odd fermion sector would be vanishing for $V_{Hd}
= {\bf 1}$.  However, the relation
$V_{Hu}^\dagger V_{Hd} = V_{\rm CKM}$ requires that if $V_{Hd} = {\bf
  1}$, then $V^\dagger_{Hu} = V_{\rm CKM}$.  If down-type mixing is
suppressed by very small off-diagonal elements, then up-type mixing is
unavoidable.

\section{Little Higgs contributions to neutral meson mixing}
\label{calculation}
We now calculate the corrections to the relevant effective
Hamiltonians in the littlest Higgs model with T-parity.  The dominant
contributions arise from box diagrams which have T-odd fermions
running within the loop, along with T-odd gauge bosons.  There are
also sub-dominant effects coming from the extended top sector of the
model, which we briefly discuss as well.

\subsection{T-odd sector contribution}

\begin{figure}[t]
\centerline{\includegraphics[width=1\hsize]{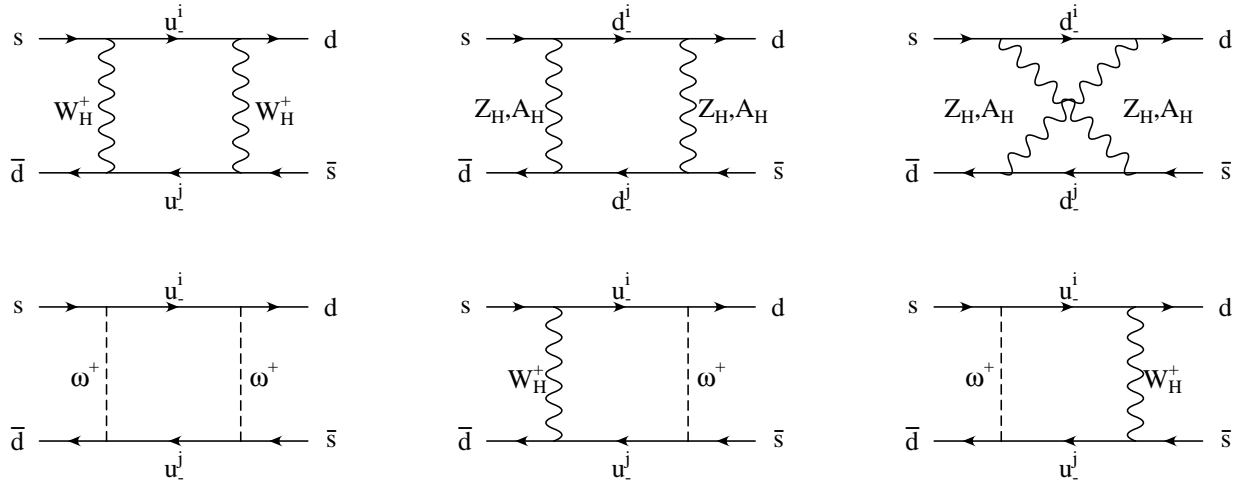}}
\caption{Box diagrams involving T-odd gauge bosons and scalars that
  contribute to particle anti-particle mixing in the littlest Higgs
  model with T-parity.  There are other diagrams, such as those with
  two neutral scalars running in the loop.  These, however, sum to
  zero.  We show only the classes of diagrams which contribute to the
  functions given in the Appendix.}
\label{fig:boxes}
\end{figure}

From the T-odd sector we get several new contributions to neutral
meson mixing. These contributions come from box diagrams that contain
heavy gauge bosons and T-odd fermions, and in general are suppressed by
a factor of $v^2/f^2$.  However, this suppression is 
vastly overcome in most regions of parameter space due to the absence of a TeV scale GIM mechanism.  The diagrams can be classified according to the
gauge boson running in the loop: $W_{H}$, $Z_{H}$, and ``mixed''
$Z_{H}$ and $A_{H}$.  These are shown in Figure~\ref{fig:boxes}.  We
have calculated these diagrams both in the 't~Hooft-Feynman and the
unitary gauge, and we now review the results.

The diagrams with internal $W_H^\pm$ and charged Goldstone bosons give
a contribution to the effective Hamiltonian which has the same functional form as the SM
calculation, with some simple replacements.
%$x_i \rightarrow y_i =
%m_f^2/m_{W_H}^2$, $V_{\rm CKM} \rightarrow V_{Hu}$ or $V_{\rm CKM}
%\rightarrow V_{Hd}$, and an overall multiplication by $v^2/4f^2$.
For the heavy neutral gauge bosons, $Z_H$ and $A_H$, we have apart from the
$W_H$-like diagrams, also ``crossed'' diagrams where the gauge bosons
attach to opposite vertices on one side of the box.  After summing over the
two types we find that each class of diagrams, namely the ``ZZ'',
``AA'', and ``ZA'' diagrams, are independently gauge invariant. We
carried out the full calculation in both 't Hooft-Feynman gauge and unitary
gauge, but gauge independence can be shown to hold in any $R_\xi$
gauge. A similar phenomenon occurs in the SM for boxes
that contain $Z_L$ and $\gamma$ \cite {Bardin:1999ak} (these diagrams
do not contribute to neutral meson mixing, of course). Furthermore,
the contribution of the diagrams containing neutral scalars vanishes after
summing over the regular and crossed diagrams. This effect can be
traced back to the fact that the coupling of the eaten T-odd Goldstone
bosons to the heavy and light
fermions is purely left handed, and that the
momentum assignment on one of the fermion lines is in the opposite
direction of fermion number flow in the crossed diagrams.

The total contribution from the T-odd sector to neutral $K$ mixing
(neglecting QCD corrections) is given by
\begin{eqnarray}
\mathcal{H}^\mathrm{odd}_\mathrm{eff} &=& \frac{G_F^2}{64 \pi^2} M_W^2
\frac{v^2}{f^2} \sum_{ij} \lambda'_i \lambda'_j \left[
F(y_i,y_j;W_H)+G(z_i,z_j;Z_H) \right. \arline
&& \left. \hspace{1.35in}+ A_1 (z_i, z_j; Z_H )+ A_2
(z_i,z_j;Z_H) \right] (\bar{s} d)_{(V-A)} (\bar{s} d)_{(V-A)}. \arline
&&\
\label{effham}
\end{eqnarray}
The functions $F$, $G$, $A_1$, and $A_2$ correspond to the contributions of the
``WW'', ``ZZ'', ``AA'', and ``ZA'' diagrams, respectively. Their
explicit form in 't~Hooft-Feynman gauge is given in the Appendix.  
%As explained above, the function $F$ is given in
%the 't Hooft-Feynman gauge, while $G$, $A_1$, and $A_2$ are gauge
%independent. 
In the above formula $y_i=m_i^2/M_{W_H}^2$ and
$z_i=m_i^2/M_{Z_H}^2$, which are identical at lowest order in $v/f$.
Ignoring the higher order effects of the $W_H - Z_H$ mass splitting,
we replace $y_i$ with $z_i$ in the rest of the text.  $m_i$ and $m_j$ are the masses of the T-odd
quarks in the loop, and the $\lambda'_i$ are
functions of $V_{Hd}$ matrix elements:
$\lambda'_i=V_{Hd}^{\,is}V_{Hd}^{\,*id}$.

As in the SM calculation, we can present the result in a more compact way. 
Imposing unitarity of $V_{Hd}$, we can re-write the effective Hamiltonian as:
\begin{equation}
\mathcal{H}^{\rm odd}_\mathrm{eff} = \frac{G_F^2}{64 \pi^2} \eta M_W^2
\frac{v^2}{f^2} \left[ \lambda'^2_3 R_2 (z_1,z_3)+2
\lambda'_2\, \lambda'_3 R_3 (z_1,z_2,z_3) + \lambda'^2_2 R_2 (z_1,z_2) \right] (\bar{s} d)_{(V-A)} (\bar{s} d)_{(V-A)},
\end{equation}
where
\begin{eqnarray}
&& R_2 (z_i, z_j) =\sum_{M \in \{ F,G,A_1,A_2 \}}  \left[M (z_i,z_i) - 2 M (z_i, z_j) + M(z_j,z_j) \right]\arline
&& R_3 (z_i, z_j, z_k) = \sum_{M \in \{ F,G,A_1,A_2 \}}  \left[M
    (z_i,z_i) - M(z_i,z_j) - M(z_i,z_k)+ M(z_j,z_k)\right], \arline
&&\ 
\label{toddcompact}
\end{eqnarray}
and $\eta$ parametrizes the effects of QCD corrections that will be
discussed in more detail below.

The effective Hamiltonians relevant to $B$ and $D$ mixing can easily
by obtained from Eq.~(\ref{toddcompact}) by simply interchanging indices in
the mixing parameters, $\lambda'_i$, and relabeling the quarks in the
four-fermion operator.

Interpreting these new contributions as shifts in physical observables
is quite easy through application of the same techniques used in the SM
calculations.  The only subtleties that arise are involved with the QCD
corrections.

Before moving on to examine the T-even contributions, it is
instructive to look at an approximate formula for the T-odd
contributions to the effective
Hamiltonian.  In particular, if we go to the limit where the T-odd
doublet spectrum is nearly degenerate, and assume that the T-odd
fermion masses are significantly larger than the T-odd gauge bosons
($\kappa \gg g$),
we find that Eq.~(\ref{effham}) reduces to the following form:
\begin{equation}
\mathcal{H}_\mathrm{eff}^{\rm odd} \approx \frac{1}{192 \pi^2 f^2}
\left[ (\delta \kappa_{12} + \delta \kappa_{23}) V^{1d}_{Hd}
  V^{*1s}_{Hd} + \delta \kappa_{23} V^{2d}_{Hd} V^{*2s}_{Hd} \right]^2
(\bar{s} d)_{(V-A)} (\bar{s} d)_{(V-A)},
\end{equation}
where $\delta \kappa_{12} = \kappa^{22}_D - \kappa^{11}_D$, and $\delta
\kappa_{23} = \kappa^{33}_D - \kappa^{22}_D$.  In this expression, it
is easy to see the GIM mechanism at work.  The lowest order terms in the
mass splitting expansion are at $\delta \kappa_{ij}^2$. 
%We note that at
%this order in $g/\kappa$, the only diagrams that contribute in
%'t~Hooft-Feynman gauge are those with charged Goldstone bosons running
%in the loop.

\subsection{Contributions from the T-even sector}

There are also contributions to flavor changing diagrams coming from
diagrams which involve the T-even partner of the top quark, $T_+$.
These have been calculated in~\cite{buras,Choudhury:2004bh,jaeyong} for the
littlest Higgs model without T-parity, and the
results from these calculations are much the same as in the littlest
Higgs model with T-parity, although some of the diagrams in that model no
longer exist due to certain couplings being forbidden by
T-parity.  For example, in diagrams with T-even fermions running in
the loop, there are no contributions which involve the heavy T-odd
gauge bosons.

\begin{figure}
\centerline{\includegraphics[width=.5\hsize]{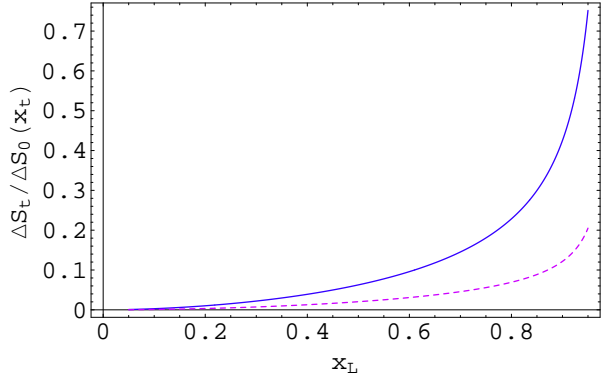}}
\caption{$\Delta S_t/S_0(x_t)$ versus $x_L$ for different values of
f, $f=1$ TeV (solid) and $f=2$ TeV (dashed). $\mathcal{O}(v^4/f^4)$
corrections are included to show the large $x_L$
behavior.}\label{fig:T-even}
\end{figure}

The leading order $\mathcal{O}(v^2/f^2)$ contribution to the
effective Hamiltonian that governs $K$ meson mixing from the
T-even sector is given by
\begin{equation}
\mathcal{H}_\mathrm{eff}^{\rm even}=\frac{G_F^2}{16 \pi^2} M_W^2
\left[ \lambda_c^2 \eta_1 \Delta S_c + \lambda_t^2 \eta_2 \Delta S_t
+ 2
  \lambda_c \lambda_t \eta_3 \Delta S_{tc} \right] (\bar{s}d)_{V_A}
  (\bar{s} d)_{V-A},
\end{equation}
where
\begin{eqnarray}
&&\Delta S_c = 0 \arline
&&\Delta S_t = -2 \frac{v^2}{f^2} x_L^2
\Big(S_0(x_t)-S_0(x_t,x_T)\Big) \arline &&\Delta S_{ct} =
-\frac{v^2}{f^2} x_L^2 \Big(S_0(x_c,x_t)-S_0(x_c,x_T)\Big),
\end{eqnarray}
and $x_T = m_{T_+}^2/m_{W_L}^2$.  The parameter $x_L$ is a function of
the couplings in Eq.~(\ref{topyuk}): 
\begin{equation}
x_L = \frac{\lambda_1^2}{\lambda_1^2+ \lambda_2^2}.
\end{equation}
These corrections arise from two effects.  First, there are explicit
new flavor changing diagrams which involve the partner of the top quark, $T_+$.
In addition, the CKM matrix is modified at order $v^2/f^2$ in the
$V_{\rm CKM}^{ti}$ elements.

To obtain these relations, we have simply taken the limit of equal
gauge couplings required by T-parity in the equations in~\cite{buras}, and
removed also the contributions from diagrams that violate T-parity.
Note that, because of those new conditions, imposing T-parity makes
the T-even contributions somewhat smaller than those in the littlest Higgs
model without T-parity.

For regions of $x_L\geq0.8$, where $\lambda_1 > \lambda_2$, we have to also
consider formally order $v^4/f^4$ contributions, which
increase linearly with $x_T$. These contributions come from box
diagrams that contain two T-even partners of the top quark. The leading
behavior of these contributions is the same as that of the littlest
Higgs model without T-parity, given by~\cite{buras}
\begin{equation}
(\Delta S_t)_{TT}\approx \frac{x_T}{4} \frac{v^4}{f^4}\,x_L^4  =
  \frac{x_t}{4} \frac{v^2}{f^2} \frac{x_L^3}{1-x_L}.
\end{equation}

In Figure~\ref{fig:T-even}, we show the ratio $\Delta S_t/S_0(x_t)$
as a function of $x_L$ at order $\mathcal{O}(v^4/f^4)$, where $S_0$
is the SM contribution.  In our analysis, we take $x_L = 0.5$, which
corresponds to the point at which the $T_+$ mass is at its minimum.
This is also the point where the contributions to the Higgs mass are
minimized.  For this `natural' value of $x_L$, these T-even contributions
are small (less than $6$\% of the SM contribution for $f=1$~TeV),
and can be neglected.

Although T-even contributions could be very large in more fine-tuned
regions of $x_L$, we note that $x_L$ cannot be arbitrarily close to 1, in
order not to violate direct search bounds on the T-odd top partner
mass, $m_{T_-}= \lambda_2 f$ (as $\lambda_1$ is increased, in order to
hold the top quark mass fixed, $\lambda_2$ must decrease, lowering the
$T_-$ mass).  In addition, we want to keep $\lambda_1$ from
entering the strong coupling regime.  We leave a study which includes
the effects of large $x_L$ for future work.\footnote{Recently, new
  little Higgs models have been constructed in which the partner of
  the top-quark is odd under T-parity~\cite{Tisodd}.  In such models, these
  contributions could vanish.  The flavor effects of the T-odd sector
  that are the primary focus of our study, however, remain unchanged with this
  modification.}

\subsection{QCD corrections}
So far the expressions we have presented did not include QCD
corrections. For the SM contributions these corrections usually suppress the
short distance predictions. For example, the
numerical values for the QCD corrections to the SM contributions are
given by
\begin{equation}
\eta_B = 0.55\pm 0.01,\ \eta_1 = 1.32\pm0.32,\ \eta_2 = 0.57\pm0.01,\ \eta_3 =
0.47\pm0.05
\end{equation}
at NLO~\cite{Buras:2005xt,Buras:1990fn,ulrich2,ulrich,ulrich3}.  A
full NLO analysis for the new physics contributions would clearly be
beyond the scope of this work, but as we will show below, we can account for
the bulk of these corrections at
leading order (LO).

For the little Higgs model with T-parity we always match onto the same
$(V-A) \otimes (V-A)$ operator. While the NLO value of the Wilson
coefficient at the high scale $\mu_H$ cannot be determined without a full one
loop calculation, the anomalous dimension will be the same as in the
SM, as it depends only on the properties of the local
operator.  This implies that we can immediately obtain 
$\eta=\alpha_s(\mu_H)^\frac{\gamma_0}{2\beta_0}$, valid at LO.

We also need to address the issue of the choice of what to take for the
high scale $\mu_H$. One might assume that $\mu_H\sim f$ is the best choice, but as we will
explain below, we choose for our study $\mu_H \sim M_W$.
\begin{itemize}
\item While the masses of $W_H$ and $Z_H$ are $gf$, the mass of $A_H$
  is $4$ times smaller: $g\,'f/\sqrt{5}$, which for $f \sim$ 1000 GeV is close
to the
top quark mass.  Therefore for diagrams that involve $A_H$ we should
use a scale lower than $f$.
\item The masses of the T-odd fermions are free parameters, so it is unclear
which scale to use when integrating them out. Furthermore, since they
couple to the gluons, their presence will lead to threshold effects
which are functions of these masses and which greatly complicate the
calculation.
\item Most importantly, the bulk of the QCD corrections result from
  running from the weak scale to the hadronic scale. Since the
  variation of $\alpha_s$ between the scales $v$ and $f$ is rather
  small, neglecting these running and threshold effects is justified, considering
  the other uncertainties. For example, in running up to $f=1000$~GeV
  from $M_W$, the effect would only reduce $\eta$ by about $8\%$.
\end{itemize}
Considering these facts, and that there are uncertainties which would
dominate these small effects, the common value for the QCD corrections
that we adopt is then:
\begin{equation}
\eta =
(\alpha_s(m_{W_L}))^\frac{\gamma_0}{2\beta_0}=(\alpha_s(m_{W_L}))^{6/23}\sim0.58\ .
\end{equation}
In order to calculate the matrix element of the resulting
effective Hamiltonian, we need to parametrize the matrix element of
the four quark operator.  This calculation is precisely the same as in
the SM as it only relies on physics at the low scale, and so the bag
parameters are identical.

\section{Results and Constraints}
\label{results}
In this section, we show our numerical bounds on the T-odd fermion
spectrum for some representative selection of textures for $V_{Hd}$.  We
first consider cases where $V_{Hu}$ and $V_{Hd}$ are diagonal up to
corrections that are of order the off-diagonal elements of $V_{{\rm
    CKM}}$.  We then analyze simple cases where the off-diagonal elements
are allowed to be large.  We find in the former cases that some small
GIM suppression is necessary to satisfy
experimental constraints.  In the large mixing scenarios, a
strong GIM suppression is necessary to avoid large contributions, and the T-odd fermion spectrum
must be nearly degenerate.

%%We neglect the mass splittings between the up and down-type T-odd
%%fermions and assume also that $M_{W_H} = M_{Z_H}$.  Including the mass
%splittings introduces only a higher order effect in the $v/f$
%%expansion.

\subsection{Near the diagonal}
The littlest Higgs model is an effective field theory valid at most to the scale $4
\pi f$.  As such, there is no reason to suspect that one particular
texture is favored over another.  However, if we begin from a basis
where the T-odd Dirac masses are diagonal, this leads to the relations $V_{Hu} = V_u$, and $V_{Hd} = V_d$.
From this relation, it is clear that, in this basis, all of the flavor and CP
violating amplitudes arise as a result of the Yukawa couplings which
give mass to the SM fermions.

Now the CKM matrix is given by $V_{{\rm CKM}} = V_u^\dagger V_d$.  From this
relation, it is clear that $V_u$ and $V_d$ cannot simultaneously be
set to the identity.  In this section, we assume that both $V_u$ and
$V_d$ are nearly equal to the identity matrix.  This is equivalent to
assuming that there is an alignment mechanism between the T-odd masses
and the SM Yukawa structure.  This assumption provides us with
a set of minimal mixing scenarios.  We take as examples two simple cases:
\begin{itemize}
\item
{\bf Case I}\hspace{.16in} $V_{Hu} = \iden$, $V_{Hd} = V_{{\rm CKM}}$
\item
{\bf Case II}\hspace{.08in} $V_{Hd} = \iden$, $V_{Hu} = V_{{\rm CKM}}^\dagger$
%\item
%{\bf Case III} $V_{Hu}^\dagger = V_{Hd}$, $V_{{\rm CKM}} =
%V_{Hu}^{\dagger 2} = V_{Hd}^2$ 
\end{itemize}

In each of these scenarios, the only parameters relevant to neutral
meson mixing are the
mass eigenvalues of the T-odd fermions.  In the first setup, the $D$ system is unaffected, and all constraints arise from neutral
$K$ and $B$ mixing.  In the second, there is no mixing in the down
type gauge and Goldstone boson interactions, and thus there are no
contributions at one loop order in the $K$ and $B$ systems.  Instead,
the $D$ system gives the only constraints.

The one feature that these scenarios both share is a relative
suppression mechanism that is borrowed from the SM CKM texture.  The
smallness of $V^{ub}_{{\rm CKM}}$ and
$V^{td}_{{\rm CKM}}$ ensure that the neutral meson mixing amplitudes will be
nearly independent of the mass of the third generation T-odd
fermions.  The constraints will primarily be on the masses of the
first two generations of T-odd fermions, because of the relatively
larger values of $V^{us}_{{\rm CKM}}$ and $V^{cd}_{{\rm CKM}}$.

\begin{figure}[t]
\centerline{\includegraphics[width=.5\hsize]{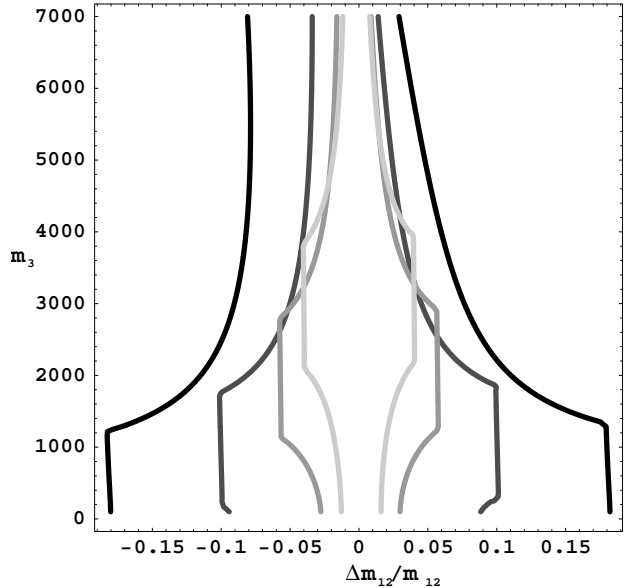}}
\caption{{\bf Case I} $f=1000$~GeV:  In these plots, $V_{Hd} = V_{{\rm
      CKM}}$.  In order of the darkest contours to lightest, the average
      mass of the first two generations varies through $m_{12} =
      500,1000,2000,3000\ {\rm GeV}$.  The $D$ system imposes no
      constraints in this scenario.}
\label{fig:caseI}
\end{figure}

In finding the bounds on the mass eigenvalues of the T-odd fermion
sector for a particular texture, we require that, for $B_d$ and $K$ mixing, the contribution from
the T-odd fermions not exceed $30\%$ of the SM contribution to the
mass splittings and $\epsilon_K$.  This is roughly when the new
physics contributions begin to exceed the long distance uncertainties
associated with the SM predictions for these observables.  We note that this process
eliminates the dependance on the bag parameters, which have rather
large theoretical uncertainties.  In the $D$ system, there is only an
experimental upper bound on the mass splitting, and the SM
short distance contribution is very small compared with this bound.
Thus, for the $D$ system we only require that
the T-odd fermion contributions not exceed this experimental upper
bound.  For every scenario, we hold the symmetry breaking scale $f$
fixed at $f = 1000$~GeV.  The contributions from new physics simply
scale as $1/f^2$, so these results can easily be extended to other
values of the breaking scale.  In each plot, the horizontal axis is
the ratio $\Delta m_{12}/m_{12} = 2 (m_2-m_1)/(m_1 + m_2)$, where
$m_{12}$ is the average mass of the first two generations, and $\Delta
m_{12}$ is the splitting $m_2 - m_1$.  On the vertical axes we plot the
dependence on the mass of the third generation T-odd quark doublet.

In Figure~\ref{fig:caseI}, we show the constraints on the mass
splitting of the first two generations of T-odd fermions as a function
of the mass of the third generation T-odd doublet.  The regions of
parameter space where the new physics contributions are smaller than
the approximate long distance uncertainties in the SM contributions lie inside the shown contours.  In this
scenario the up-type CKM, $V_{Hu}$, is diagonal, and thus $D^0 - \bar{D}^0$
mixing has no contributions from new physics.  The features in the
plot that cause the narrowing of the internal regions as $m_3$ varies away from $m_{12}$ are due to the influence of the
$\epsilon_K$ observable.  If the CP violating
phase $\delta_{13}^d$ is set to zero, the contours are nearly vertical.

\begin{figure}[t!]
\centerline{\includegraphics[width=.5\hsize]{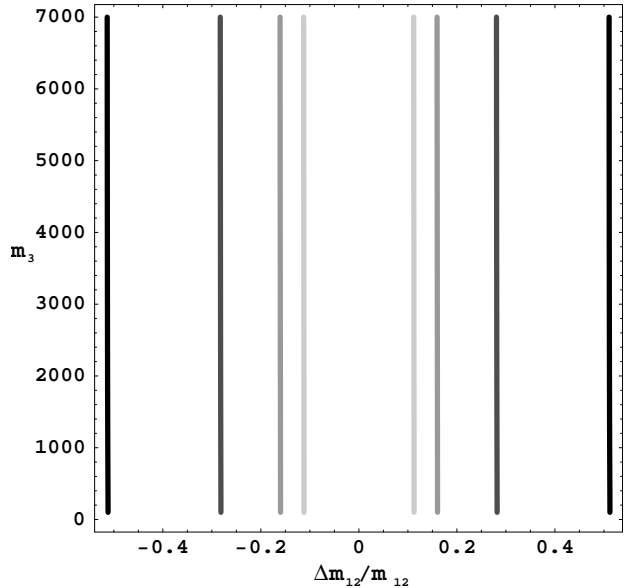}}
\caption{{\bf Case II} $f=1000$~GeV:  In this plot, we show the
  results for the case where $V_{Hd} = \iden$.  Again, from darkest to
  lightest, the average mass of the first two generations increases as
  $m_{12} = 500,1000,2000,3000\ {\rm GeV}$.  In this scenario, only
  the $D$ system is affected, and so the contours correspond to the
  points at which the T-odd fermion contributions exceed the
  experimental upper bounds.}
\label{fig:caseII}
\end{figure}
In Figure~\ref{fig:caseII}, we set instead the down-type Yukawa
interactions to be diagonal.  As mentioned, the constraints in this
region come only from the $D$ system mass splitting.  There are essentially no
constraints on the mass of the third generation T-odd doublet.  The degeneracy
required in the first two generations is quite relaxed, now
varying between $50$ and $10$\% as the average mass $m_{12}$ is
increased.  We note, however, that this would change as the experimental
bounds on the $D$ meson mass splitting are improved.  For example, if
the bound on the mass splitting comes down by a factor of ten, the
required degeneracy between the first two generations of T-odd
fermions then varies between about $16$ and $4$\% as
$m_{12}$ varies between $500$ and $3000$~GeV.

\subsection{Going away from the diagonal}
As mentioned, another possibility for the textures is to have large
off diagonal elements in $V_{Hd}$.  There is no reason to assume that
there is an alignment between the T-odd mass textures and the SM
Yukawa couplings.  We note that this requires that there are also,
simultaneously, large off diagonal elements in $V_{Hu}$ which must
cancel in the relation $V_{Hu}^\dagger V_{Hd} = V_u^\dagger V_H
V_H^\dagger V_d =  V_{{\rm CKM}}$.  This is
easy to realize in a natural way if most of this mixing comes in
through the T-odd Yukawa textures.

In this section, we study the corrections that arise
when the angles $s_{ij}^d$ in Eq.~(\ref{newCKM}) are taken to be large.  In these cases, we find
that not only is a degeneracy required between the first two
generations, but the entire flavor spectrum of the T-odd vector-like
quarks must often be degenerate.  We consider four scenarios:
\begin{itemize}
\item
{\bf Case IIIa} $s_{13}^d = 0.5$, $\delta_{13}^d = \delta_{13}^{SM}$, $s_{ij}^d = s_{ij}^{SM}$ otherwise.
\item
{\bf Case IIIb} $s_{13}^d = 0.5$, $\delta_{13}^d = 0$, $s_{ij}^d = s_{ij}^{SM}$ otherwise
\item
{\bf Case IVa} $s_{13}^d = 0.5$, $s_{12}^d = 0.7$, $s_{23}^d = 0.4$,
$\delta_{13}^d = \delta_{13}^{SM}$
\item
{\bf Case IVb} $s_{13}^d = 0.5$, $s_{12}^d = 0.7$, $s_{23}^d = 0.4$,
$\delta_{13}^d = 0$
\end{itemize}
In cases IIIa and IIIb, we allow one of the angles to be large.  We pick
specifically $s_{13}^d$ to be large, as it is this angle to which the
third generation mass dependence is sensitive.  As $\epsilon_K$ is
a strong factor in the analysis, we look at the case where it receives no
contributions by setting $\delta_{13}^d$ to zero, relegating all new CP
violation to the up-type quark interactions.  In cases Va and Vb, we
chose some order one values for the three mixing angles, and
again look at cases where the new CP violating phase is either all in the
down-type, or all in the up-type quark interactions.

\begin{figure}[t!]
\centerline{\includegraphics[width=.5\hsize]{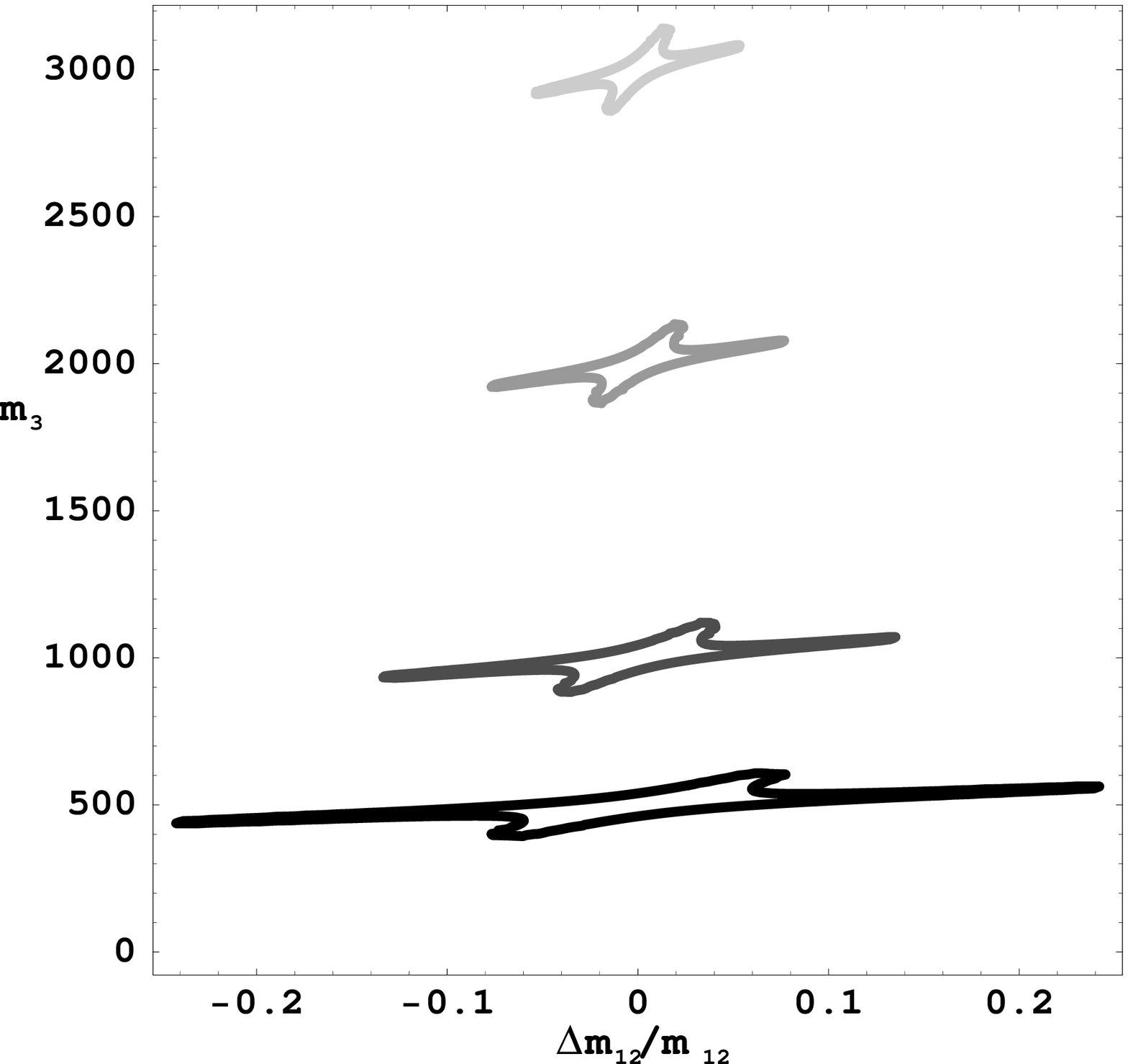}\ \includegraphics[width=.5\hsize]{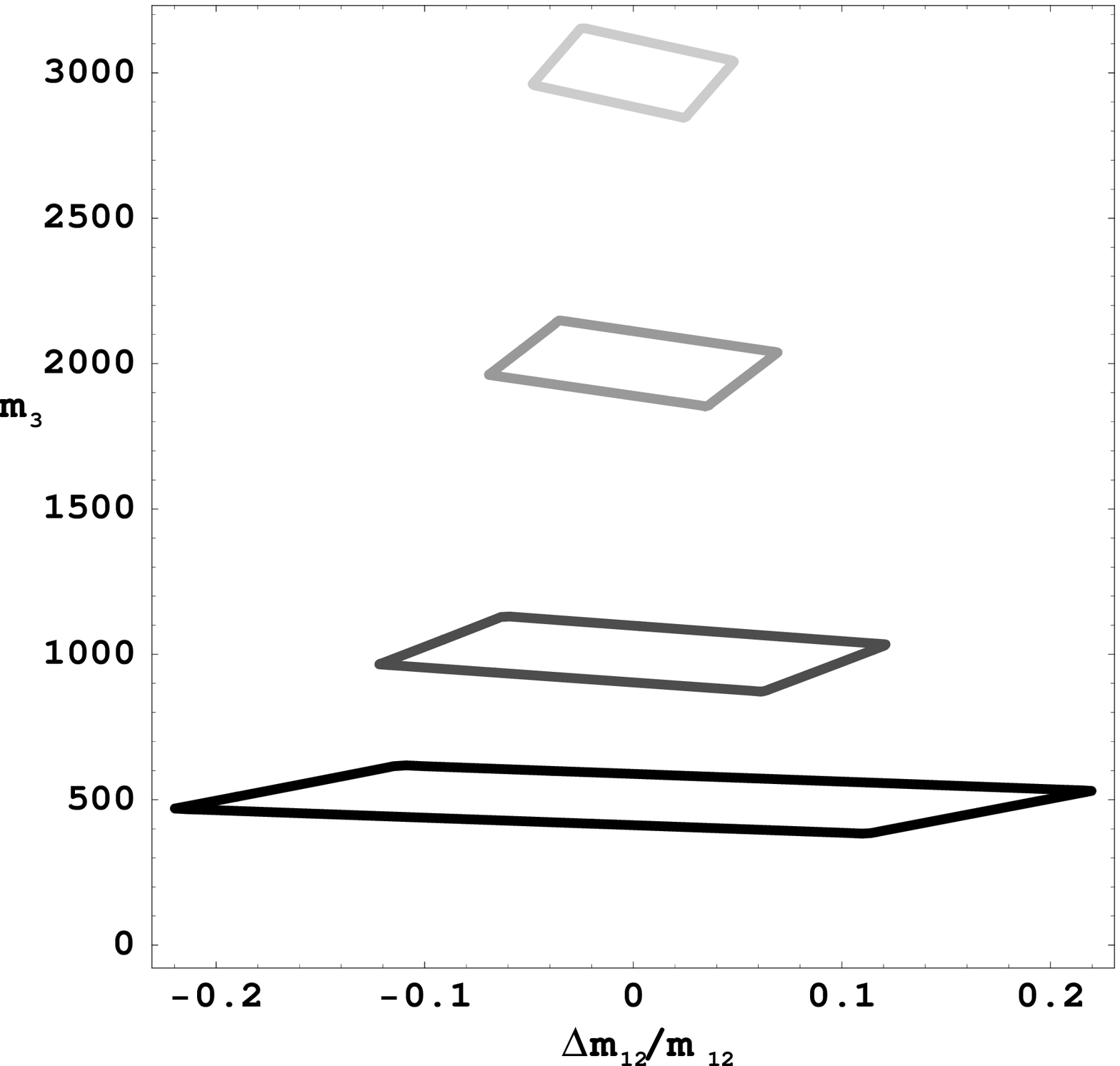}}
\caption{{\bf Case III} $f=1000$~GeV:  In this plot, the angle
  $s_{13}^d = 0.5$, while the other angles are equal to the SM CKM
  angles.  The contributions from new physics are generally much
  larger, and thus a
  stronger GIM suppression is necessary.  The contours, from darkest
  to lightest, are for $m_{12} = 500,1000,2000,3000$~GeV.  In the plot
  on the left, $\delta_{13}^d = \delta_{13}^{SM} = 1.05$, while on the
  right this phase is set to zero.}
\label{fig:bigs13H}
\end{figure}

In Figure~\ref{fig:bigs13H}, we show the constraints on the masses in
this case where $s_{13}^d$ is large.  A large $s_{13}^d$ implies order
one contributions to $V_{Hd}^{3d}$ and $V_{Hd}^{1b}$.  It is clear from this
figure that a
degeneracy is now required in all three generations of T-odd fermions.  For a
generic choice of order one mixing angles, it is expected that such a
universally degenerate spectrum is required.  We show the results when
the CP violating phase is set both to the SM value,
$\delta_{13}^d = 1.05$, and to $\delta_{13}^d = 0$.  The dramatic
difference between these two types of scenarios indicates the severe
sensitivity of the $\epsilon_K$ observable to new flavor physics.  

\begin{figure}[t!]
\centerline{\includegraphics[width=.5\hsize]{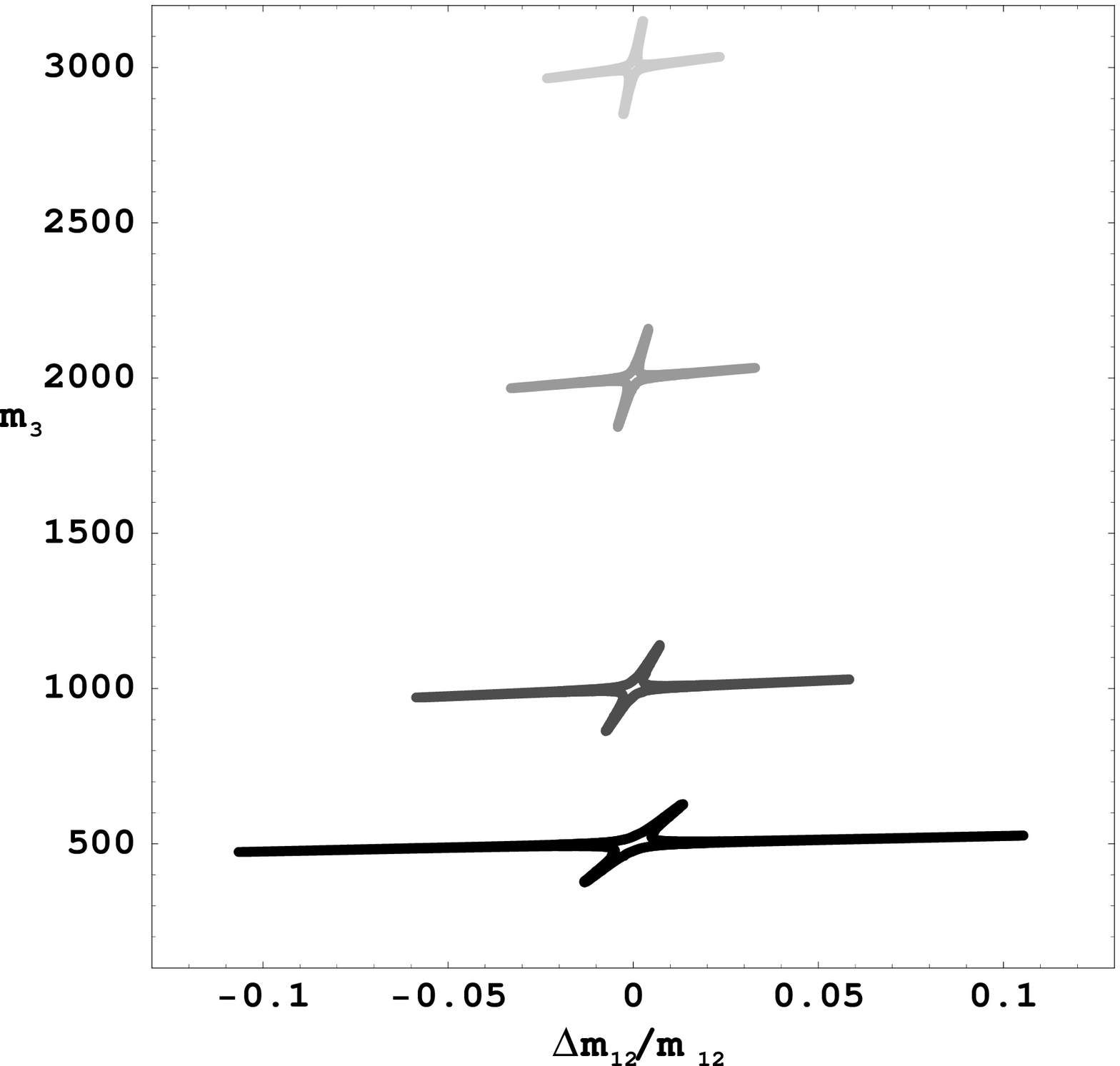}\ \includegraphics[width=.5\hsize]{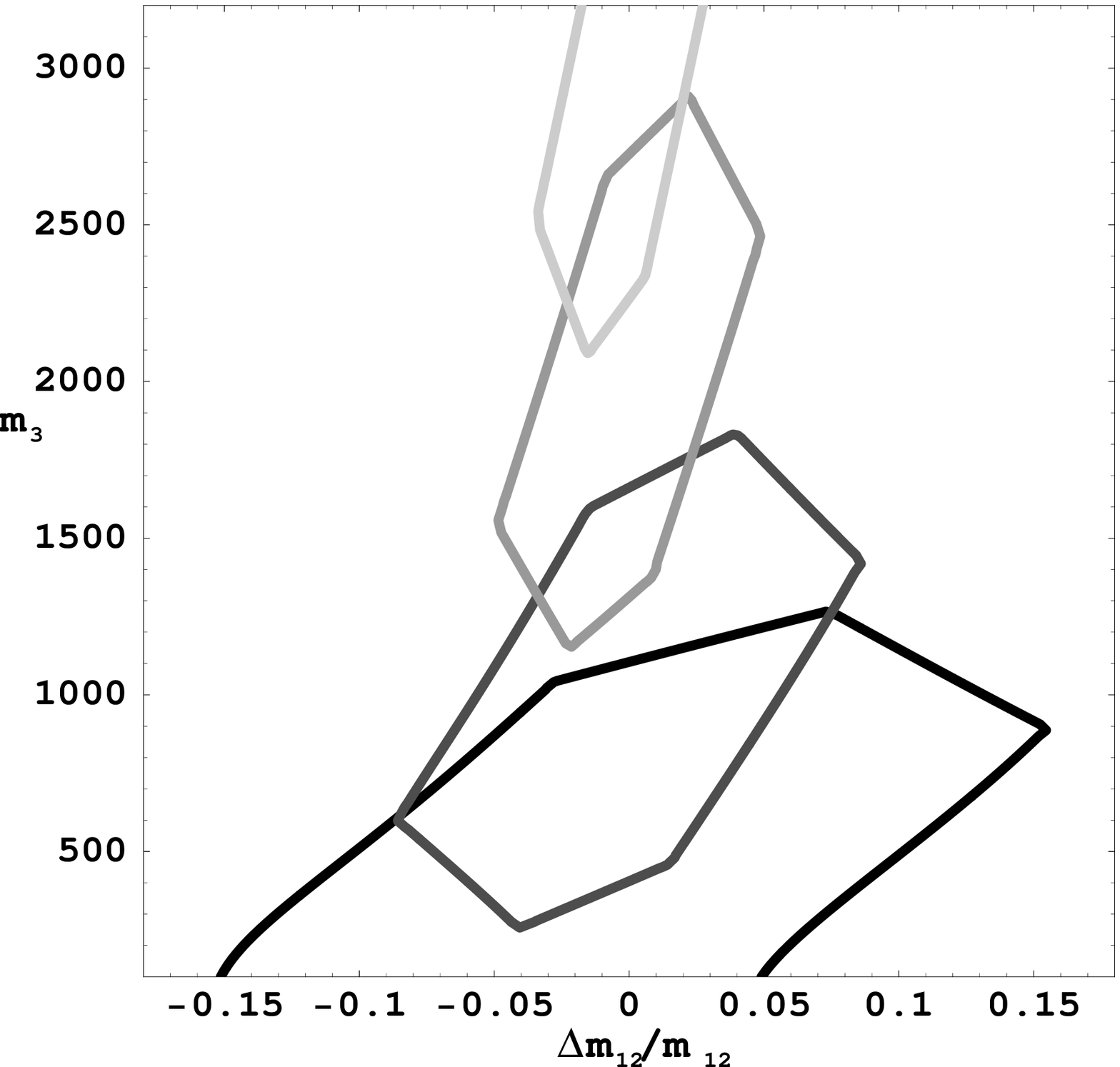}}
\caption{{\bf Case IV} $f=1000$~GeV: In this plot, all of the angles
  are taken to be somewhat large:  $s_{12}^d = 0.5,\ s_{13}^d = 0.7,\
  s_{23}^d = 0.4$.  The contours, from darkest
  to lightest, are for $m_{12} = 500,1000,2000,3000$~GeV.  In the plot
  on the left, $\delta_{13}^d = \delta_{13}^{SM} = 1.05$, while on the
  right this phase is set to zero.  In the plot on the right, the required
  degeneracy in the third generation is relaxed.}
\label{fig:bigall}
\end{figure}

In Figure~\ref{fig:bigall}, we take all of the mixing angles to be
somewhat large.  We find that this scenario is far more constrained
then all the others if the phase $\delta_{13}^d = \delta_{13}^{SM}$.
There are some narrow windows where degeneracies of up to $10$\%
are allowed, but the majority of the parameter space where corrections
are small is in the $1$\% range.  However, when the angle $\delta_{13}^d$
is taken to be small, it happens that there is a cancellation in
$V_{Hd}^{3d}$, such that the third generation mass $m_3$ is relatively
unconstrained.

\subsection{$B_s$ mixing}

\begin{figure}[t]
\centerline{\includegraphics[width=.5\hsize]{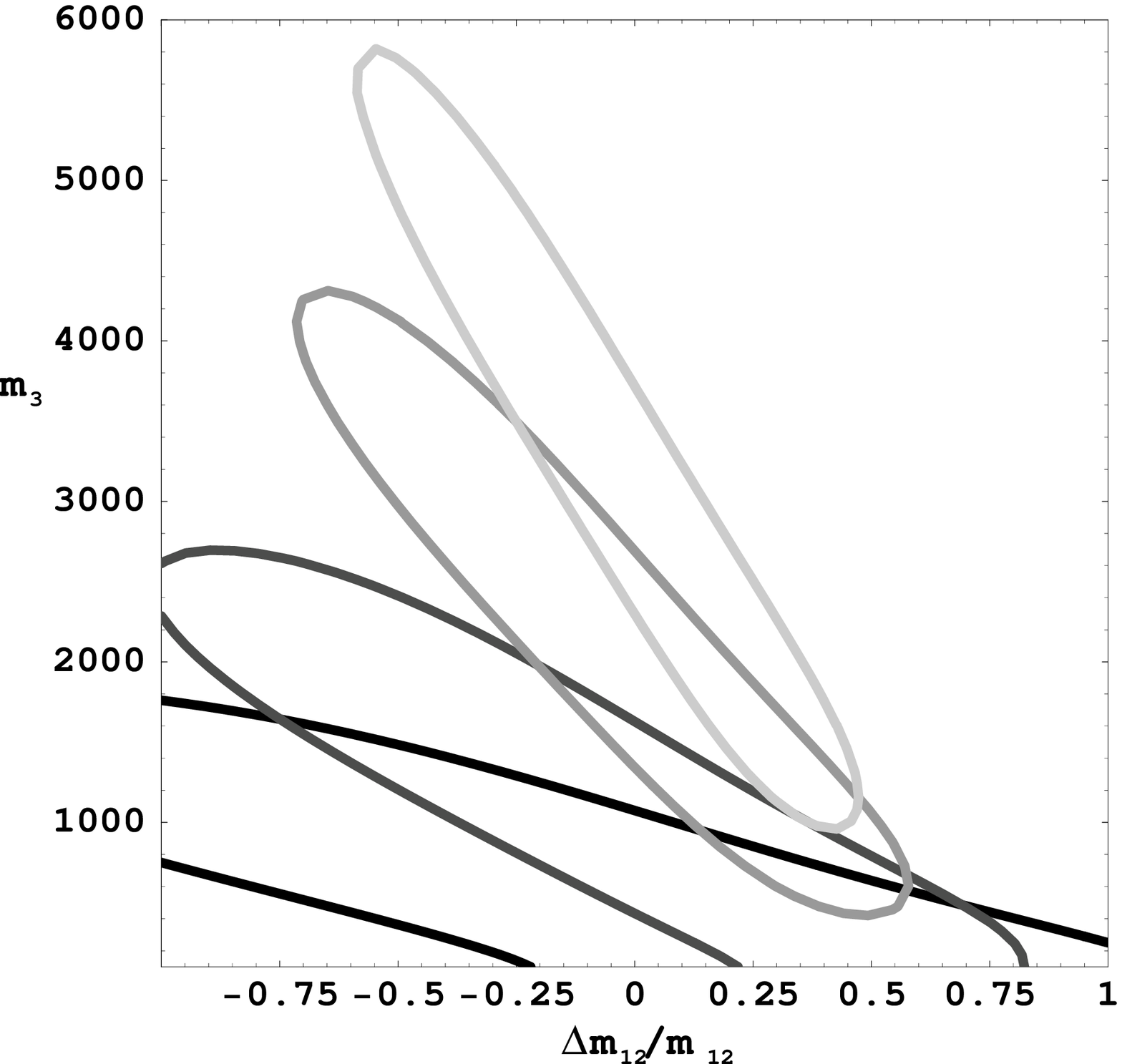}\ \includegraphics[width=.5\hsize]{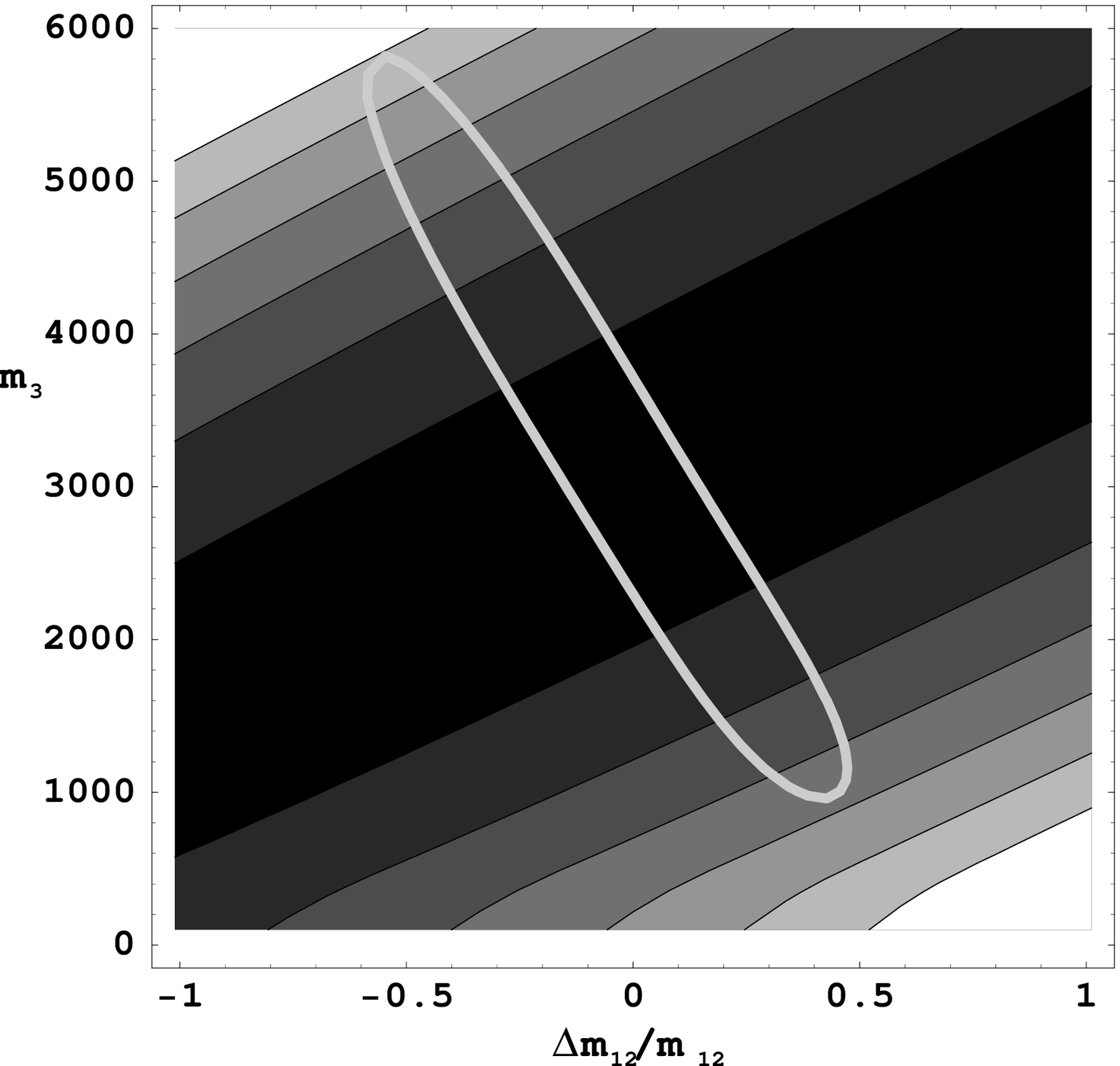}}
\caption{{\bf Case V} $f=1000$~GeV:  In these plots, the angle
  $s_{23}^d = 1/\sqrt{2}$, while the other angles are set to zero.  In the
  plot on the left, the contours, from darkest
  to lightest, are for $m_{12} = 500,1000,2000,3000$~GeV, and show the constraints from the $K$, $B_d$, and $D$
  systems.  In the plot on the right, we overlap the constraint
  contour for $m_{12} = 3000$~GeV with a plot of the
  enhancement in the $B_s$ mass splitting relative to the SM
  contribution.  From darkest to
  lightest, this enhancement is a factor of $2,4,6,8,10,$ and $12$,
  respectively.}
\label{fig:Bsmix}
\end{figure}

Of all the scenarios that we have considered so far, $B_s$ mixing is not
strongly affected if the T-odd fermion spectrum is
constrained such that the new physics contributions to $K$, $B_d$, and
$D$ mixing do not exceed the bounds that we impose.  However,
there may be some special choices of textures that we have
not considered that only strongly modify the $B_s$ system.  It is well
known that this can occur in supersymmetry~\cite{MNGROSS}.  With this in mind,
we have identified a texture that does not significantly affect $K$
and $B_d$ mixing, but which is able to enhance $B_s$ mixing.  A simple set of angles that
achieves this is
\begin{itemize}
\item
{\bf Case V} $s_{23}^d = 1/\sqrt{2}$, $s_{12}^d=0$, $s_{13}^d=0$, $\delta_{13}^d = 0$.
\end{itemize}
The constraints from the other neutral meson systems are very weak
here.  It is primarily the $D$ system which restricts the allowed
parameter space.  In this scenario, by varying the T-odd fermion masses within
the allowed contours, the mass splitting in the $B_s$
system can be enhanced by as much as a factor of $12$.\footnote{We are
  especially grateful to Matthias Neubert for suggesting that such a
  scenario is possible.}  The
constraints, along with a plot where we show the enhancement of the
$B_s$ mass splitting for fixed $m_{12} = 3000$~GeV are shown in
Figure~\ref{fig:Bsmix}.  It is interesting to note that the degeneracy
required in the first two generations of T-odd fermions is more or less
completely relaxed.  We emphasize
that we have not performed an exhaustive search, and it is possible that there
are other textures where the allowed enhancement of $B_s$ mixing is
even larger.

%In summary, we found that the results of our analysis for the T-even
%sector is very similar to that of the littlest Higgs without
%T-parity presented in~\cite{buras}. 
%For example, we found that there
%is no correction to the Fermi constant $G_F$ at order
%$\mathcal{O}(v^2/f^2)$, which is usually calculated from the tree
%level muon decay
%In particular, the numerical values of all the CKM elements
%not involving the top quark do not change significantly.  Thus, we
%would not consider further details of T-even sector in our
%presentation.

\section{Conclusions}
\label{conclusions}

Little Higgs models with T-parity necessarily introduce new mirror
fermions in order to cut off UV sensitive contributions to
four-fermion contact operators that are constrained primarily by
studies at LEP.  These fermions introduce a new flavor
structure to the model, and lead to new tree level flavor changing
currents involving SM fermions and mirror fermions.  We have done a
first exploratory study of this flavor structure, and found constraints on
the mirror fermion mass spectrum from a one loop analysis of neutral
meson mixing.  We have noted that it is not possible to adjust all of
the new flavor structure to be completely diagonal, due to relations with
the CKM mixing already present in the SM.

For order one mixing parameters, we find that the mirror
fermion mass spectrum must be degenerate to within a few percent or less.
If the new mixing parameters are taken to be small, then this is
significantly relaxed.  In particular, if all mixing is relegated to
the up-type quark interactions, only $D$ mixing is affected, and a
degeneracy of only $50$\% or so is required between the first
two generations of T-odd fermions.  We note that improved experimental constraints on the $D$ meson mass splitting could significantly
restrict such scenarios.  We have found that the
$\epsilon_K$ observable plays a significant role in the fits if there
is a CP violating phase in $V_{Hd}$.  

%In the regions of parameter space where neutral meson mass splitting
%bounds are satisfied, we have also calculated the
%predicted values of $B_s$ mixing, which could be measured in the near
%future.  We find, for the textures considered, that $B_s$ mixing is
%shifted only slightly from its SM prediction if the constraints on $B_d$,
%$K$, and $D$ mixing are satisfied.

We have also studied the $B_s$ system, identifying a scenario in
which $B^0_s - \bar{B}^0_s$ mixing can differ substantially from the
SM prediction while still satisfying constraints on the other neutral
meson mixing observables.  In the setup we have considered, the
enhancement of the mass splitting can be as large as a factor of
$12$.  Such scenarios are of particular interest for experimental
studies of the $B_s$ system.  Also, in this scenario, the constraints
on the first two generations are much more relaxed than in the others
considered.

We wish to make clear that little Higgs models with T-parity are not ruled out
in any way by this study.  This analysis should instead serve as a
guide to what properties any UV completion of this structure
should have.  This is in close analogy with studies of the supersymmetric flavor problem, which have been an
essential tool in constructing mechanisms of supersymmetry breaking which are
consistent with low energy phenomenology.

We note that this is only an introduction to the flavor physics of
this model.  There are many other observables which are sensitive to
this flavor structure, such as rare decays and lepton flavor violating
processes.  Including rare decay
processes in an analysis would possibly require a closer degeneracy in
the mass spectrum, although this needs to be checked.  In addition, we
have assumed that the SM CKM fit is unchanged, when in fact additional
contributions to observables (especially $\epsilon_K$) can
change the best fit values of the CKM elements.  A full global
analysis would remedy this situation, however many more observables
must be computed and included to render such a fit meaningful.

\section*{Acknowledgments}
We would like to thank Enrico Lunghi, Patrick Meade, Matthias Neubert,
and Maxim Perelstein for helpful discussions and suggestions during
the preparation of this manuscript.  JH is supported by the
U.S. Department of Energy under grant DE-AC02-76CH03000.  SJL and GP
are supported in part by the National Science Foundation under grant
PHY-0355005.

\section*{Appendix}

The gauge and Yukawa interactions of the littlest Higgs model with
T-parity lead to tree level flavor changing currents which can, at one
loop, affect SM observables such as neutral meson mixing.  After
identifying the mass eigenstates, the Lagrangian can be expanded,
leading to the relevant Feynman rules.  These rules are given in
Table~\ref{flavorrules!}.  While the conjugate interactions are not
shown explicitly, they are easily derived.  One should note that the
Yukawa type interactions with the eaten Goldstone bosons do not have an $i$
prefactor.  Because of this, the associated conjugate Feynman rules have an
additional minus sign.
\begin{table}[h]
\center{\begin{tabular}{|l|c||l|c|}
\hline 
Particles & Vertices & Particles & Vertices \\

\feyn
{\bar{u}_{Hi} W_H^{+\mu} d^j}
{ i \frac{g}{\sqrt{2}} (V_{Hd})^i_j \gamma^\mu P_L}
{\bar{d}_{Hi} W_H^{-\mu} u^j}
{ i \frac{g}{\sqrt{2}} (V_{Hu})^i_j \gamma^\mu P_L}

\feyn
{\bar{u}_{Hi} Z_H^\mu u^j}
{i \frac{g}{2} (V_{Hu})^i_j \gamma^\mu P_L}
{\bar{d}_{Hi} Z_H^\mu d^j}
{-i \frac{g}{2} (V_{Hd})^i_j \gamma^\mu  P_L}

\feyn
{\bar{u}_{Hi} A_H^\mu u^j}
{-i \frac{g'}{10} (V_{Hu})^i_j \gamma^\mu P_L}
{\bar{d}_{Hi} A_H^\mu d^j}
{-i \frac{g'}{10} (V_{Hd})^i_j \gamma^\mu P_L}

\feyn
{\bar{u}_{Hi} d^j \omega^+}
{-\frac{1}{\sqrt{2} f} M_{u_H^i} (V_{Hd})^i_j P_L}
{\bar{d}_{Hi} u^j \omega^-}
{-\frac{1}{\sqrt{2} f} M_{d_H^i} (V_{Hu})^i_j P_L}

\feyn
{\bar{u}_{Hi} u^j w_3}
{-\frac{1}{2 f} M_{u_H^i} (V_{Hu})^i_j P_L}
{\bar{d}_{Hi} \omega_3 d^j}
{\frac{1}{2 f} M_{d_H^i} (V_{Hd})^i_j P_L}

\feyn
{\bar{u}_{Hi} u^j \eta}
{\frac{1}{\sqrt{20} f} M_{u_H^i} (V_{Hu})^i_j P_L}
{\bar{d}_{Hi} d^j \eta}
{\frac{1}{\sqrt{20} f} M_{d_H^i} (V_{Hd})^i_j P_L}

\feyn
{\bar{\nu}_{Hi} W_H^{+\mu} e^j}
{ i \frac{g}{\sqrt{2}} (V_{He})^i_j \gamma^\mu P_L}
{\bar{e}_{Hi} W_H^{-\mu} \nu^j}
{ i \frac{g}{\sqrt{2}} (V_{H\nu})^i_j \gamma^\mu P_L}

\feyn
{\bar{\nu}_{Hi} Z_H^\mu \nu^j}
{i \frac{g}{2} (V_{H\nu})^i_j \gamma^\mu P_L}
{\bar{e}_{Hi} Z_H^\mu e^j}
{-i \frac{g}{2} (V_{He})^i_j \gamma^\mu  P_L}

\feyn
{\bar{\nu}_{Hi} A_H^\mu \nu^j}
{i \frac{g'}{10} (V_{H\nu})^i_j \gamma^\mu P_L}
{\bar{e}_{Hi} A_H^\mu e^j}
{i \frac{g'}{10} (V_{He})^i_j \gamma^\mu P_L}

\feyn
{\bar{\nu}_{Hi} e^j \omega^+}
{-\frac{1}{\sqrt{2} f} M_{\nu_H^i} (V_{He})^i_j P_L}
{\bar{e}_{Hi} \nu^j \omega^-}
{-\frac{1}{\sqrt{2} f} M_{e_H^i} (V_{H\nu})^i_j P_L}

\feyn
{\bar{\nu}_{Hi} \nu^j w_3}
{-\frac{1}{2 f} M_{\nu_H^i} (V_{H\nu})^i_j P_L}
{\bar{e}_{Hi} \omega_3 e^j}
{\frac{1}{2 f} M_{e_H^i} (V_{He})^i_j P_L}

\feyn
{\bar{\nu}_{Hi} \nu^j \eta}
{\frac{1}{\sqrt{20} f} M_{\nu_H^i} (V_{H\nu})^i_j P_L}
{\bar{e}_{Hi} e^j \eta}
{\frac{1}{\sqrt{20} f} M_{e_H^i} (V_{He})^i_j P_L}
\hline
\end{tabular}}
\caption{This table contains the Feynman rules relevant to flavor
  changing physics.  The
  conjugate interactions are not included, but can easily be derived
  from the listed expressions.}
\label{flavorrules!}
\end{table}

The functions resulting from evaluation of the box diagrams are given by
\begin{eqnarray}
&&F(y_i,y_j;W_H) = \frac{1}{(1-y_i)(1-y_j)} \left(1-\frac{7}{4} y_i
y_j\right) +\frac{y_i^2 \log y_i}{(y_i - y_j) (1-y_i)^2} \left( 1- 2
y_j + \frac{y_i y_j}{4} \right) \arline
&& \hspace{3.35in} -\frac{y_j^2 \log y_j}{(y_i - y_j) (1-y_j)^2} \left( 1- 2
y_i + \frac{y_i y_j}{4} \right) \arline
&&G(z_i,z_j;Z_H) = -\frac{3}{4} \left[\frac{1}{(1-z_i)(1-z_j)}+\frac{z_i^2 \log
    z_i}{(z_i - z_j) (1-z_i)^2} - \frac{z_j^2 \log z_j}{(z_i - z_j)
    (1-z_j)^2} \right] \arline
&&A_1 (z_i, z_j ; Z_H ) = -\frac{3}{100 a} \left[ \frac{1}{(1- z'_i)(1-z'_j)} + \frac{z'_i z_i \log 
    z'_i}{(z_i-z_j) (1-z'_i)^2} - \frac{z'_j z_j \log
    z'_j}{(z_i-z_j) (1-z'_j)^2} \right] \arline
&&A_2 (z_i, z_j ; Z_H ) = -\frac{3}{10} \left[ \frac{\log a}{(a-1) (1- z'_i)(1-z'_j)} + \frac{z_i^2 \log
    z_i}{(z_i-z_j) (1-z_i) (1- z'_i)} \right. \arline
&& \hspace{3.25in} \left. - \frac{z_j^2 \log
    z_j}{(z_i-z_j) (1-z_j)(1-z'_j)} \right], \arline
&& \ 
\end{eqnarray}
where $a = M_{Z_H}^2/M_{A_H}^2 \approx 5/\tan^2 \theta_w$, and $z'_i = a z_i$.  The function $F$
contains the contributions from the charged T-odd scalars and gauge
bosons, while $G$ contains the contributions involving two $Z_H$
propagators.  $A_1$ contains the contributions
from diagrams with two $A_H$ propagators, while $A_2$ contains the
contributions from diagrams with both a $Z_H$ and an $A_H$ propagator
running in the loop.

We note that in unitary gauge the expression for the $F$ function is
not the same, and in fact contains divergent terms.  These
cancel when the sum over flavors running in the box diagrams is
performed, and unitarity of the mixing matrices is imposed.  It is
only after this summation that the calculations in the two different
gauges can be compared.  In contrast, the $G$, $A_1$, and $A_2$
functions which correspond to neutral current contributions are
already gauge invariant.

\end{document}